\documentclass[review,3p]{elsarticle}%review 
\usepackage[T1]{fontenc}
\usepackage[utf8]{inputenc} 								
\usepackage[english]{babel}						
\addto\captionsenglish{}

\usepackage{epsfig}
\usepackage{subfig}
\usepackage[dvipsnames]{xcolor}
\usepackage[colorlinks]{hyperref}

\usepackage{amsthm}
\usepackage{amssymb}
\usepackage{amsmath}
\usepackage{mathtools}
\usepackage{textcomp}

\usepackage{lineno}

\journal{Advances in Water Resources}

\usepackage{bm}
\newcommand{\W}{\bm{W}}
\newcommand{\Mcw}{\mathtt{M}_\text{cw}}
\newcommand{\Mcs}{\mathtt{M}_\text{cs}}
\newcommand{\Mq}{\mathtt{M}_\text{q}}
\newcommand{\Mm}{\mathtt{M}}
\newcommand{\MS}{\mathtt{MS}}
\newcommand{\M}{\bm{\mathcal{M}}}
\newcommand{\A}{\bm{\mathcal{A}}}
\newcommand{\Fr}{\mathtt{Fr}}

\newcommand{\Uacc}{\mathtt{U{acc}}}
\newcommand{\NUacc}{\mathtt{NU{acc}}}
\newcommand{\U}{\mathtt{L}}
\newcommand{\Tol}{\mathtt{Tol}}
\newcommand{\Sp}{\mathtt{Sp}}
\newcommand{\Eg}{\mathtt{E}_z}
\newcommand{\MF}{\mathtt{MF}}
\newcommand{\UC}{\bm{U}}     			% val
\newcommand{\LL}{\bm{L}}				% val
\newcommand{\RR}{\bm{R}}				% val

\usepackage{booktabs}       								% per le linee nelle tabelle
\usepackage{siunitx}												% per le tabelle di soli numeri
\usepackage{multirow}												% per creare celle multiriga nelle tabelle

\usepackage{changes}
\definechangesauthor[name={Francesco}, color=ForestGreen]{F}
\definecolor{bblue}{rgb}{0.23,0.4,0.7}
\definechangesauthor[name={Valerio}, color=bblue]{val}
\definechangesauthor[name={Nunzio}, color=red]{NUN}
\setremarkmarkup{[#2]}
\usepackage{cancel}

\begin{document}

\begin{frontmatter}



\title{Mathematical study of linear morphodynamic acceleration and derivation of the MASSPEED approach}


\author[unife]{F. Carraro}
\author[eth]{D. Vanzo}
\author[unife]{V. Caleffi}
\author[unife]{A. Valiani}
\author[eth]{A. Siviglia}

\address[unife]{Department of Engineering, University of Ferrara, Italy}
\address[eth]{Swiss Federal Institute of Technology, Laboratory of Hydraulics, Hydrology and Glaciology, Z\"urich, Switzerland}

\begin{abstract}
Morphological accelerators, such as the MORFAC (MORphological acceleration FACtor) approach \citep{Roelvink2006}, are widely adopted techniques for the acceleration of the bed evolution, which reduces the computational cost of morphodynamic numerical simulations. 
In this work we apply a non-uniform acceleration to the one-dimensional morphodynamic problem described by the de Saint Venant-Exner model by multiplying all the spatial derivatives by an individual constant ($\geq 1$) acceleration factor.
The final goal is to identify  the best combination of the three accelerating factors for which i) the bed responds linearly to hydrodynamic changes; ii) a consistent decrease of the computational cost is obtained.
The sought combination is obtained by studying the behaviour of an approximate solution of the three eigenvalues associated with the flux matrix of the accelerated system.
This approach allows to derive a new linear morphodynamic acceleration technique, the MASSPEED (MASs equations SPEEDup) approach, and the \textit{a priori} determination of the highest acceleration allowed for a given simulation. 
In this new approach both mass conservation equations (water and sediment) are accelerated by the same factor, differently from the MORFAC approach where only the sediment mass equation is modified.
The analysis shows that the MASSPEED gives a larger validity range for linear acceleration and requires smaller computational costs than that of the classical MORFAC approach.  
The MASSPEED approach is implemented within an example code, using an adaptive approach that applies the maximum linear acceleration similarly to the Courant–Friedrichs–Lewy stability condition.
Finally, numerical simulations have been performed in order to assess accuracy and efficiency of the new approach.
Results obtained in the long-term propagation  of a sediment hump demonstrate the advantages of the new approach. 
\end{abstract}


\end{frontmatter}

%\linenumbers

\section{Introduction}
\label{sec:intro}
Reducing the computational costs of numerical simulations of the morphological evolution in rivers, estuaries and coastal areas is a critical issue for engineers and geomorphologists \cite[e.g.][]{Coco:2013,Roelvink2016,Siviglia2016}. 
Even though simulation tools of physical systems have greatly benefited from the increasing computational power over the last decades thanks to progress in CPU performances and parallelization technologies, the use of morphodynamic upscaling techniques is still widely popular and becomes essential when long-term evolutions must be predicted \cite[e.g.][]{Coco:2013,Roelvink2016,Siviglia2016}.
Classical approaches for morphodynamic acceleration have been developed primarily for costal and estautuarine applications  \cite{Devriende1993,latteux1995techniques,Roelvink2006}.
Among them, the MORFAC (MORphological acceleration FACtor) approach \cite{Lesser2004,Roelvink2006} has been introduced 
in the context of coastal applications, with the purpose of efficiently describing the overall morphodynamic effect of a high number of repeated tides.
The MORFAC approach is now standard in state-of-the-art comercially available numerical morphological codes \cite{Ranasinghe2011}.
It is daily employed by engineers for solving practical problems in costal and estaurine environments \cite{Roelvink2016,Coco:2013} and increasingly used for the simulation of river morphodynamics \cite[e.g.][]{duro2016numerical,mosselman2016five,Nicholas2013,Oorschot2015}.
The key idea of the MORFAC approach is to accelerate the morphological evolution increasing the time bottom variations by a given constant ($>1$) factor, thus accelerating the morphodynamic processes.
This is effectively obtained by multiplying the sediment flux in the Exner equation by a constant ($>1$) acceleration factor updating the bed and flow within the same time step.
According to \citet{Ranasinghe2011} this approach can be adopted only if the morphological response to the hydrodynamic forcing is linear during one morphological time step. 
Therefore, the key issue in the application of such an approach is to find the maximum acceleration factor (critical MORFAC in the literature) that can be applied in the numerical simulation.
One of the first attempts to assess the accuracy and stability of the MORFAC approach has been carried out by \citet{Roelvink2006}, who also performed a comparison among different acceleration techniques.
Even though the recent advancements \citep{li2010,Ranasinghe2011} to develop a theoretical background to detect this value are significant, the critical MORFAC is still often set by trial-and-error procedures \cite[e.g.][]{duro2016numerical,mosselman2016five,Nicholas2013,Oorschot2015}. 

The main goal of this paper is to  develop a  robust theoretical background for the development of linear morphodynamic acceleration techniques.
This will be obtained by performing a mathematical study to quantitatively identifying the limit of application for given hydraulic and sediment-transport conditions, thereby overcoming the limits of a trial and error approach.
In this work we focus on the one-dimensional morphodynamic problem described by the de Saint Venant-Exner (dSVE) system of equations.
The proposed mathematical framework for studying linear morphodynamic accelerations is developed by considering a non-uniformly accelerated dSVE model, i.e. each of the three governing equations (water and sediment continuity, and conservation of momentum) is accelerated by constant acceleration factors ($>1$), namely $\Mcw$, $\Mcs$ and $\Mq$. 
The final goal is to identify  the most convenient combination of the three accelerating factors, to be applied  within each single time step,  for which i) the bed still responds linearly to hydrodynamic changes; and ii) a consistent decrease of the computational time is obtained.
The analysis is carried out by studying an approximate solution for the eigenvalues of the flux matrix associated with the accelerated system, taking advantage of the typically small value attained by the ratio  between the sediment and the water discharge.
This analysis is an extension of that performed by \citet{Lyn1987} for the standard de Saint Venant-Exner (non-accelerated) system.
Our aims is that, from the standard MORFAC approach obtained by setting $\Mcw=\Mq=1,\Mcs>1$, to seek the optimal combinations of these.
The analysis of the eigenvalues of the non-uniformly accelerated system of the dSVE allows also derivation of a new linear morphodynamic acceleration technique, which has been dubbed MASSPEED (MASs equations SPEEDup).
This is obtained by setting $\Mcw=\Mcs>1,\,\Mq=1$ and is characterized by a larger validity range for linear acceleration and higher computational speed-up to that of the classical MORFAC approach. 
Finally, the new MASSPEED approach is implemented using an adaptive approach, similarly to that used for implementing the Courant–Friedrichs–Lewy numerical stability condition.
It is then applied to the long-term propagation of a sediment hump with the aim of demonstrating that it is able i) to correctly predict the time evolution with and without friction terms; ii) to decrease considerably the computational costs.

The paper is structured as follows: \S \ref{sec:MathModel} briefly reviews the governing equations and their main mathematical properties.
In \S \ref{sec:morph_up} we present the new general framework for morphodynamic acceleration and in \S \ref{sec:LmaT} we derive the MASSPEED approach.
In \S \ref{sec:Non_Lin_An} we introduce two numerical strategies to compute the maximum acceleration factors.
In \S \ref{sec:application} we present numerical results for the long term propagation of a sediment hump assessing the advantages of the new approach proposed in this paper.
Conclusions are drawn in \S \ref{sec:conclusion}.

\section{Analysis of the morphodynamic mathematical model}
\label{sec:MathModel}
We consider a one-dimensional morphodynamic model, which describes the flow over an erodible bottom. The bed is
assumed to be composed of uniform sediments which are transported by the flow as bedload. 

\subsection{Governing equations}
\label{sec:matmod}
The governing equations are obtained under shallow water
conditions imposing mass conservation for the fluid and solid
phases and  momentum conservation. In  one-dimension they read
\begin{equation}
	\left\{
	\begin{aligned}
		&\frac{\partial h}{\partial t} + \frac{\partial q}{\partial x} = 0\,, \\
		&\frac{\partial q}{\partial t} + \frac{\partial}{\partial x}\left(\frac{1}{2}gh^2 + \frac{q^2}{h}\right) + gh\frac{\partial z}{\partial x}=  -gh\, s_f\,,\\
		&\frac{\partial z}{\partial t} + \xi\frac{\partial q_s}{\partial x} = 0\,,
	\end{aligned}
	\right.
	\label{eq:def_SW-Ex}
\end{equation}
where $x$ is the longitudinal coordinate, $t$ is time, $h(x,t)$ and $z(x,t)$ denote the water depth and the bottom elevation respectively, $q(x,t)$ and $ q_s(x,t)$ indicate the liquid and bed-load discharge per unit width, $g$ is the gravitational acceleration, $s_f$ is the friction slope and $\xi=1/(1-p)$, where $p$ is the porosity of the riverbed (hereafter we assume $p=0.4$).

The governing system \eqref{eq:def_SW-Ex} is composed by three partial differential equations (PDEs) in five unknowns, namely $h(x,t)$, $q(x,t)$, $z(x,t)$, $s_f(x,t)$ and $q_s(x,t)$. Therefore, two extra relations are required to close the system. The friction  term is provided by a classical closure, namely
\begin{equation}
s_f=\frac{q^2}{K^2_{s}h^{10/3}}\,,
\end{equation} 
where $K_s$ is the Strickler coefficient.
For the sake of simplicity, the bed-load flux is assumed to be a power function of the velocity \citep{Grass1981},
\begin{equation}
	q_s = A_g \, u^m \,,
\label{eq:Grass}
\end{equation}
where $A_g$ and $m$ are two constant parameters. Hereafter we assume $m=3$.  

\subsection{Eigenvalues and characteristic curves}
The system of governing equations (\ref{eq:def_SW-Ex}) (\textit{original system}) can be cast in quasi-linear form as
\begin{equation}
	\frac{\partial\W}{\partial t}+\A(\W)\frac{\partial\W}{\partial x} = \bm{S}(\W)\,,
	\label{eq:qLin}
\end{equation}
where $\W$ is the vector of the conservative variables, $\A(\W)$ is the flux matrix and $\bm{S}(\W)$ is the vector of the source terms. 
It follows from Eq.~\eqref{eq:def_SW-Ex} that
\begin{equation}
	\W= \begin{bmatrix} h \\	q \\ z \end{bmatrix}\,,  \quad
	\A(\W)=
	\begin{bmatrix}
								0 												&								1														&			0	  		\\
							c^2-u^2											&	 						 2u														&		 c^2			\\
			\xi\frac{\partial q_s}{\partial h} 	&	 		\xi\frac{\partial q_s}{\partial q}		&			0
	\end{bmatrix}\,, \quad
	\bm{S}(\W)= \begin{bmatrix} 0 \\	-c^2 s_f \\ 0 \end{bmatrix}\,,
	\label{eq:def_WAS}
\end{equation}
where
$u=q/h$ is the depth averaged velocity and
$c=\sqrt{gh}$ is the propagation celerity of gravitational waves.
Using Eq.~\eqref{eq:Grass}, the two terms on the last row of matrix $\A(\W)$ can be written as
\begin{equation}
\psi = \xi \frac{\partial q_s}{\partial q} \quad\text{and}\quad \xi \frac{\partial q_s}{\partial h} = -u\psi  \quad\text{with}\quad  \psi = 3\, \xi\left( \dfrac{q_s}{q}\right) = 3\,\xi \, \left(A_g\,g\, \Fr^2\right)\,,
\label{eq:phi_grass}
\end{equation}
where $\psi$ is the transport parameter, depending on the ratio between the flux of the sediments and the water discharge, while $\Fr=u/c$ is the Froude number.
The parameter $\psi$ is usually small ($\psi \ll 1$), at least in the common range of river and coastal typical sediment transport rates.
Given these definitions, the flux matrix $\A(\W)$ can be rewritten as
 \begin{equation}
	\A(\W)=
	\begin{bmatrix}
								0 												&								1														&			0	  		\\
							u^2\left(\dfrac{1}{\Fr^2}-1\right)											&	 						 2u														&		 \dfrac{u^2}{\Fr^2}		\\
			-u \psi 	&	 		\psi		&			0
	\end{bmatrix}\qquad.
	\label{eq:Jacobian}
\end{equation}

The characteristic polynomial and the eigenvalues of \eqref{eq:Jacobian} are obtained by setting $|\A -\lambda \bm{I}|=0$, where $\bm{I}$ is the 3$\times$3-identity matrix; the characteristic polynomial reads
\begin{equation} \label{eq_char_pol}
\lambda^3-2  u \lambda^2 
+ (\Fr^2 - \psi-1) \frac{u^2}{\Fr^2}  \lambda +  \frac{u^3}{\Fr^2}\psi = 0\;.
\end{equation}
If a power law formula for the solid transport is used, as that adopted in Eq.~\eqref{eq:Grass}, the three eigenvalues $\lambda_1, \lambda_2, \lambda_3$, the solutions of the cubic polynomial \eqref{eq_char_pol}, are always real, therefore the governing system is always hyperbolic \cite{Cordier2011}.
For more general sediment transport formulas, where the non dimensional solid discharge is expressed as a function of the dimensionless Shields parameter (e.g., \citet{Meyer-Peter1948}), if the friction term is closed using a Manning approach, the governing system is hyperbolic provided that $\Fr<6$ \cite{Cordier2011}.
This latter condition is usually satisfied under \emph{natural} conditions.

\begin{figure}[tb]%
\centering
\includegraphics[width=1\columnwidth]{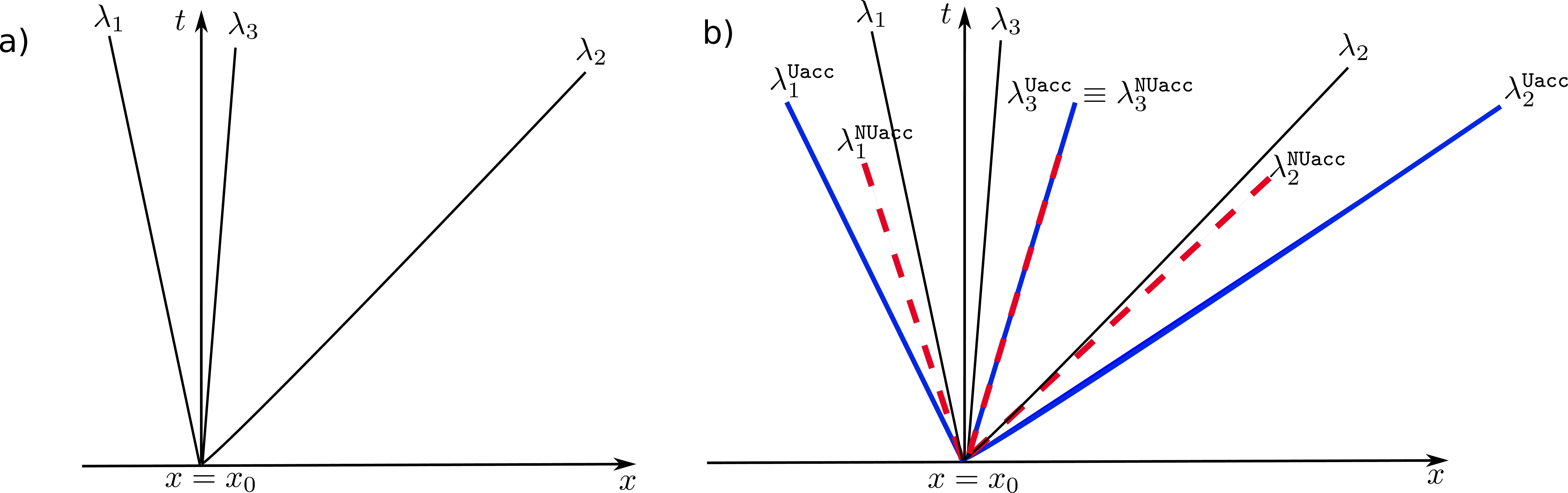}%
\caption{Sketch of (linearized) characteristic curves in the phase space. In panel a) the typical situation for subcritical conditions ($u>0$) is given. In panel b) the original system (black thin lines) the uniformly accelerated system ($\Uacc$) (blue thick lines) and the  non-uniformly accelerated system ($\NUacc$) (red dashed lines) are sketched. The definitions of uniformly and non-uniformly accelerated systems are given in the text.}%
\label{fig:accEig}%
\end{figure}

On the ($x-t$) plane (phase space) the eigenvalues are celerities associated to characteristic curves along which small disturbances propagate.
Herein, the term celerity is applied broadly to mean the velocity of propagation of a disturbance, either on the water surface or on the bed.
For very small disturbances the characteristic curves can be approximated by straight lines.
The situation for a small disturbance generated at time $t=0$ in $x_0$, in case of subcritical flow ($\Fr<1$) is given in Figure \ref{fig:accEig}a.

\subsection{Weak interaction between hydrodynamics and morphodynamics: the decoupled system}
\label{sec:approx_sol}

Though the three roots of the characteristic polynomial \eqref{eq_char_pol} can be determined exactly, the derivation of an approximate solution, obtained by a perturbative analysis, is useful for interpretation. Typically, the transport parameter $\psi$ may be estimated as small as $O(10^{-3}$--$10^{-5})$ \cite[e.g.][]{Lanzoni2006,li2010,Lyn1987,Lyn2002} therefore it seems reasonable to expand $\lambda$ in powers of $\psi$ as follows:
\begin{equation} \label{eq:expans}
\lambda = \lambda^{(0)} + \psi\, \lambda^{(1)} + \psi^2\, \lambda^{(2)} \;.
\end{equation}
We then substitute (\ref{eq:expans}) into Eq.~\eqref{eq_char_pol}, equate likewise powers of $\psi$ and look for the approximate solution of the three eigenvalues $\lambda_i \; (i=1,2,3)$.
At the leading order $O(\psi^0)$, a classical result is found:
one of the eigenvalues ($\lambda_3$) vanishes and the remaining two reduce to those found in the fixed bed case ($\lambda_\mathtt{H1,H2}$), namely
\begin{equation} \label{eq_roots_leading}
\lambda_\mathtt{H1,H2} \equiv \lambda^{(0)}_{1,2} \cong \lambda_{1,2} = \left[\left(\Fr \mp 1\right)+O (\psi)\right] c\;. 
\end{equation}  
At the next order $O(\psi)$  small ``morphodynamic'' corrections for the two hydrodynamic eigenvalues, $\lambda_{1,2}$,  are found and the  third eigenvalue, associated with  bed level changes, arises.
Writing also the second order term,  $\lambda_3$ reads as
\begin{equation} \label{eq_morpho_root_OPsi}
\lambda_{3} \equiv \lambda_{b} =  \left[\left(\frac{\Fr}{1-\Fr^2}\right) \psi -\left[ \frac{\Fr\left(\Fr^2+1\right)}{\left(1-\Fr^2\right)^3}\right] \psi^2  +  O (\psi^3)\right]c\;.
\end{equation}  
Eq.~\eqref{eq_morpho_root_OPsi} clearly shows that the present perturbative analysis is valid provided that $\Fr$ is small, in fact when $\Fr\rightarrow1$, $\lambda_{3}$ tends to infinity. Moreover,  by comparing  Eq.~\eqref{eq_roots_leading} with Eq.~\eqref{eq_morpho_root_OPsi} it is seen that the relative order of magnitude of the celerities associated with the characteristic curves of the hyperbolic system is rather different \cite[e.g.][]{Devries1965}. 

The behavior of such curves is well known: far from the critical state (i.e. $\Fr\ll1$) the celerity of a small amplitude bed wave  is considerably smaller than that of small amplitude hydrodynamic waves \cite{Lyn2002}.
Therefore, the bed interacts only weakly with the water surface, thus justifying an approach in which the equations governing hydrodynamics are solved separately from those governing  morphodynamics. 
It follows that, from a mathematical point of view, the problem can be described separately by the Saint-Venant equations for the hydrodynamics and by a simple nonlinear wave equation for morphodynamics,
\begin{equation}\label{eq:wave_eq}
\frac{\partial z}{\partial t} + \lambda_b \frac{\partial z}{\partial x} = 0 \;.
\end{equation} 
Taking into account Eq.~\eqref{eq_roots_leading} and Eq.~\eqref{eq_morpho_root_OPsi}, neglecting the friction term and writing the mass and momentum balance laws for the fluid phase in terms of characteristic variables $u_1$ and $u_2$ \cite{Toro2009}, the complete system of governing equations, which approximates system (\ref{eq:def_SW-Ex}) when $\psi \ll 1$, can be written in \textit{decoupled form} as
\begin{equation}
	\begin{aligned}
		&\frac{\partial }{\partial t}\begin{bmatrix} u_1 \\	u_2  \\ z \end{bmatrix} + \Lambda \frac{\partial }{\partial x} \begin{bmatrix} u_1 \\	u_2 \\ z  \end{bmatrix}  = 0\;,
	\end{aligned}
	\label{eq:def_SW-Ex_decoupled}
\end{equation}
where 
\begin{equation}
	\Lambda=
	\begin{bmatrix}
\lambda_\mathtt{H1} & 0&0 \\
0 &	 \lambda_\mathtt{H2}  &0 \\
0 &	0   & \lambda_b \\
	\end{bmatrix}=
	\begin{bmatrix}
(\Fr -1)+O(\psi) &0 \\
0 &	(\Fr+1)+O(\psi)  &0 \\
0 &	0   & \dfrac{\Fr}{1-\Fr^2}\psi +O(\psi^2) \\
	\end{bmatrix} c\;.
	\label{eq:def_lambda_H}
\end{equation} 
According to Eqs.~(\ref{eq:def_SW-Ex_decoupled}), in the phase space ($x-t$ plane), small hydrodynamic perturbations propagate along the characteristic curves $\frac{dx}{dt}=\lambda_\mathtt{H1,H2}$ while small bed perturbations travels along the curve $\frac{dx}{dt}=\lambda_{b}$.
From Eqs.~(\ref{eq:def_SW-Ex_decoupled}) emerges that, if $\psi \ll 1$, $z$ is a characteristic variable of the morphodynamic problem. In \ref{app:Val} a linearized example is proposed to validate the decoupled form of the dSVE system.

\section{A general framework for morphodynamic acceleration}
\label{sec:morph_up}
Small hydrodynamic and bed disturbances propagate along the characteristic curves with celerities given by the eigenvalues of the matrix \eqref{eq:Jacobian}.
Therefore, linear acceleration of the propagation of small disturbances can be obtained by increasing the slope of such characteristic curves.
If we consider the system of governing equations (\ref{eq:qLin}) and neglect friction terms ($\bm{S}=\bm{0}$),  acceleration can be obtained by multiplying from the left the \textit{original} flux matrix by the acceleration matrix $\M$, namely
\begin{equation}
\frac{\partial \W}{\partial t}+\M\,\A\frac{\partial \W}{\partial x} = 0 \,,\qquad \M = \begin{bmatrix}
					\Mcw	&	0	& 0 		\\
					0	&	\Mq		&	0		\\
					0	&	0  		&	\Mcs
			\end{bmatrix},
\label{eq:NUaccSystem}
\end{equation}
in which we consider three scalar acceleration coefficients,
$\Mcw$ for the water continuity equation,
$\Mq$ for the momentum equation,
and $\Mcs$ for the sediment continuity equation.

In this section two configurations of $\M$ are discussed: a simple case of uniform acceleration $(\Uacc)$ of the whole system with $\Mcw=\Mq=\Mcs=\Mm>1$, and the case of non-uniform acceleration $(\NUacc)$ in which each equation is accelerated by a specific, and in general different, factor ($\Mcw\geq1,$  $\Mq\geq1,$ $\Mcs\geq1$).
In particular, for the non-uniformly accelerated system we present an approximate set of eigenvalues and suitable definitions for the numerical speed-up.

\subsection[The trivial case of uniform acceleration (Uacc)]{The trivial case of uniform acceleration $(\Uacc)$}
\label{sec:morph_Uacc}

In this section, a \emph{uniformly accelerated} system is analyzed from a computational point of view. We set $\Mcw=\Mq=\Mcs=\Mm>1$ in\eqref{eq:NUaccSystem} and the resulting system of equations reduce to:
\begin{equation}
	\frac{\partial \W}{\partial t}+\Mm \, \A \frac{\partial \W}{\partial x} = 0 \;.
\label{eq:UaccLinSyst}
\end{equation}
Making use of the eigenvalues and eigenvectors properties, it is easy to verify that all the three eigenvalues of $\A$ scale linearly with $\Mm$.
In other words, we have the uniformly accelerated eigenvalues $\lambda_i^{\Uacc} = \Mm\,\lambda_i$.
In this case, hydrodynamic and morphodynamic information are accelerated linearly by the same factor; therefore, the slopes of the three characteristic curves will be larger and all proportional to $\Mm$.
The corresponding situation in the phase space is depicted in Fig.~\ref{fig:accEig}b, where characteristic curves related to the uniformly accelerated system are displayed with thick-blue continuous lines.

It is interesting to analyse the consequences of this acceleration when a numerical solution of system \eqref{eq:UaccLinSyst} is sought and the adopted scheme must satisfy the Courant-Friedrichs-Lewy (CFL) condition.
Here we recall only that the CFL condition is necessary to obtain that the numerical solution (using a finite volume or finite difference method) is stable and converge to the exact solution as the grid is refined \cite{Toro2009}.

If the spatial domain is discretized with a grid having a constant mesh size $\Delta x$, the numerical solution of the original system (\ref{eq:qLin}) is advanced in time by a time step $\Delta t$ that must satisfy the following CFL condition:
\begin{equation}
	\Delta t \le \mathtt{CFL}\,\frac{\Delta x}{\max(\lambda_i)}.
	\label{eq:CFL}
\end{equation}
where: $\mathtt{CFL}$ is the Courant–Friedrichs–Lewy number with specific values depending upon the selected time integration technique; $\max(\lambda_i)$ is the maximum eigenvalue associated to the flux matrix (\ref{eq:Jacobian}).
If we consider $u>0$, $\max(\lambda_i)=\lambda_2$, as in Fig.~\ref{fig:accEig}a.
For the uniformly accelerated system the maximum time step that can be adopted is 
\begin{equation}
	\Delta t_\Uacc = \mathtt{CFL}\,\frac{\Delta x}{\lambda_2^{\Uacc}}\,.
	\label{eq:AccCFL}
\end{equation}
Since $\lambda_2^{\Uacc}=\Mm\,\lambda_2$, Eqs.~\eqref{eq:CFL} and \eqref{eq:AccCFL} imply
\begin{equation}
	\Delta t_\Uacc = \mathtt{CFL}\,\frac{\Delta x}{\Mm\,\lambda_2}\quad\Rightarrow\quad\Delta t_\Uacc=\frac{\Delta t}{\Mm} \;.
\label{eq:Dt_Uacc}
\end{equation}
This means that, if a uniform acceleration $\Mm$ is imposed, time integration of the accelerated system requires a time step which is $\Mm$-times smaller to that of the original system. 
Moreover, to take into account the acceleration of the morphological evolution, the time must be scaled in the accelerated framework.
In particular, if we consider a given propagation time $t_p$ related to the simulation performed by using the original system, the corresponding propagation time related to the uniformly accelerated system is $t_p/\Mm$.

Therefore, the propagation time and the time step in the accelerated framework are both scaled by $\Mm$, so that the numerical solution at a given output time $t_p$ of the original and the uniformly accelerated systems requires the same number of time steps; \emph{consequently no computational gain is obtained by using the $\Uacc$ procedure}. 

\subsection[The case of non-uniform acceleration (NUacc)]{The case of non-uniform acceleration $(\NUacc)$}
\label{sec:morph_NUacc}

Given the conclusion of the previous section, if a computational gain is sought, a non-uniform acceleration must be considered.
The final goal is to find an appropriate combination of the three accelerating factors ($\Mcw\ge1, \Mq\ge1, \Mcs\ge1$)  with a twofold objective, as follows:
\begin{itemize}
\item [(i)] \textit{obtaining a linear acceleration for morphodynamics}. This can be obtained when the bottom time evolution can be described by the following  accelerated wave equation 
\begin{equation}\label{eq:acc_wave_eq}
\frac{\partial z}{\partial t} + \Mcs \lambda_b \frac{\partial z}{\partial x} = 0 \;.
\end{equation}
Therefore, we look for the physical and mathematical conditions under which the accelerated system of governing equations can be written in the decoupled form (\ref{eq:def_SW-Ex_decoupled}). The linear acceleration allows to precisely describe the propagation of the bed level in the accelerated phase space, providing a linear correspondence between the time scales of the non-accelerated and the accelerated morphodynamic process.   
\item [(ii)] \textit{increasing the numerical speed-up}. This means that, if the morphodynamic process is accelerated by a constant factor $\Mcs$, the corresponding acceleration of the hydrodynamic process should not reduce the numerical speed-up.  
That is, the largest eigenvalue of the system of the governing equations ($\lambda_2$) must be accelerated by a factor smaller than the morphodynamic acceleration factor $\Mcs$. 
\end{itemize}

Graphical representation on the phase space of the two conditions i) and ii) is given in Fig.~\ref{fig:accEig}b. 

\subsubsection[Approximate eigenvalues for the (NUacc) system]{Approximate eigenvalues for the $\NUacc$ system}
\label{sec:approx_sol_NU}
We derive an approximate solution of the three eigenvalues of the $\NUacc$ system by taking advantage of the typically small magnitude of $\psi$ and adopting a perturbative analysis similar to that carried out in \S \ref{sec:approx_sol}.
This solution is used in \S \ref{sec:LmaT} to find the conditions under which a particular choice of [$\Mcw,\,\Mq,\,\Mcs$] can satisfy the requirement i).

The characteristic polynomial associated to system \eqref{eq:NUaccSystem} is obtained by imposing $|\M\,\A-\lambda \bm{I}|=0$; it reads
\begin{equation} \label{eq_char_pol_Uacc}
\lambda^3-2 \Mq u \lambda^2 
+ \Mq (\Fr^2 \Mcw-\Mcs \psi-\Mcw) \frac{u^2}{\Fr^2}  \lambda +  \Mcs  \Mcw \Mq \frac{u^3}{\Fr^2}\psi = 0\;,
\end{equation}
compare this with \eqref{eq_char_pol}.
 
Assuming expansion (\ref{eq:expans}) for the solution of equation (\ref{eq_char_pol_Uacc}) to $O(\psi)$ we obtain
\begin{equation} \label{eq_roots_0}
\frac{\lambda_{1,2}^{\NUacc}}{c} = \Mq \left[\Fr \mp \sqrt{\Fr^2 + \frac{\Mcw}{\Mq}\left(1- \Fr^2\right)}\right]  +  
\Mcs \left[\dfrac{\sqrt{\dfrac{\Mcw}{\Mq}\dfrac{1-\Fr^2}{\Fr^2}+1}+\left(1-\dfrac{\Mcw}{\Mq}\right)}{\sqrt{\dfrac{\Mcw}{\Mq}\dfrac{1-\Fr^2}{\Fr^2}+1}+\left(1-\dfrac{\Mcw}{\Mq}\right)+\frac{1}{\Fr^2}\dfrac{\Mcw}{\Mq}} \right]\frac{\psi}{2\Fr}\;.
\end{equation}  
At $O(\psi^2)$, the celerity associated with the bed level changes  is given by
\begin{equation} \label{eq_roots_2}
\frac{\lambda_3^{\NUacc}}{c} = \Mcs \left[ \left(\frac{\Fr}{1-\Fr^2}\right) \psi -\left(\frac{\Mcs}{\Mcw} \frac{\Fr\left(\Fr^2+1\right)}{\left(1-\Fr^2\right)^3}\right) \psi^2  \right]\;.
\end{equation}  
It is clear that, also for the non-uniformly accelerated system, when $\Fr \rightarrow 1$, $\lambda_3^{\NUacc} \rightarrow \infty$.
Therefore the assumptions on the perturbation expansion are no longer valid \cite{Lyn1987}.

\subsubsection{Theoretical and computational speed-up}
\label{sec:def_SpSpCPU}
The goal of this section is to derive a general theoretical definition for the speed-up of the non-uniform acceleration approach.
This is the mean which allows us to quantify the computational gain for a given morphological simulation.
We demonstrated that a uniform acceleration applied to both hydrodynamic and morphological evolutions does not result in any computational speed-up.
Quite contrary, we expect to have a speed-up when the eigenvalue corresponding to the bottom evolution, $\lambda^{\NUacc}_3$, is accelerated by a factor $\Mcs$ compared to the corresponding eigenvalue of the original system $\lambda^{\NUacc}_3=\Mcs \lambda_3$ and at the same time the maximum eigenvalue of the hydrodynamic system,  $\lambda_2^{\NUacc}$, results to be accelerated by a factor $<\Mcs$ as compared to its homologous in the original system $\lambda_2$.
For the sake of generality, we define the averaged theoretical speed-up over a simulation time $T_\mathtt{S}$ as the ratio between the instantaneous morphological $\mathtt{R_M}(\tau)$  and hydrodynamic $\mathtt{R_H}(\tau)$ acceleration,
\begin{equation}
	\Sp =\frac{1}{T_\mathtt{S}}\,\int\displaylimits_0^{T_\mathtt{S}}{ \frac{\mathtt{R_M}(\tau)}{\mathtt{R_H}(\tau)} \,d\tau}\;.
\label{eq:def_Sp}
\end{equation}
If the condition for Eq.~\eqref{eq:acc_wave_eq} is satisfied, i.e. the $\NUacc$ system can be written in a decoupled way, the acceleration terms $\mathtt{R_{M,H}}(\tau)$ are the ratio between the accelerated and original eigenvalues,
\begin{equation}
	\mathtt{R_{M}(\tau)} = \frac{\lambda_3^{\NUacc}(\tau)}{\lambda_3(\tau)}\,, \quad \mathtt{R_{H}(\tau)} = \frac{\lambda_2^{\NUacc}(\tau)}{\lambda_2(\tau)}. 
\label{eq:def_RHRM}
\end{equation}
It is worth noting that, if the acceleration factors $\mathtt{R_{M,H}}$ are set constant over the whole simulation time, equation \eqref{eq:def_Sp} simplifies to $\Sp = \mathtt{R_{M}}/\mathtt{R_{H}}$.

In a similar way we can define the computational speed-up as the ratio between the computational time (CPU time) of the reference solution and the accelerated one,
\begin{equation}
\Sp^\mathtt{CPU} = \frac{\mathtt{CPU(Ref.Sol.)}}{\mathtt{CPU(Acc.Sol.)}} \,.
\label{eq:def_numSp}
\end{equation}

\section{Linear morphodynamic acceleration techniques}
\label{sec:LmaT}
In this section we consider the classical MORFAC ($\MF$) and the newly proposed MASSPEED ($\MS$) acceleration techniques.
For both methods, the conditions under which a linear acceleration is possible are identified, and the maximum theoretical speed-up is quantified.
Moreover, in \ref{app:Val} a simple application is introduced to explain why the expression of the exact acceleration of $\lambda_3^\NUacc$ is a necessary condition to correctly reproduce the riverbed profile with both MORFAC and MASSPEED approaches.

\subsection{The classical MORFAC approach}\label{sec:clsMORFAC}
%--------------------------------------------
\begin{figure}[t]
\includegraphics[scale=0.5]{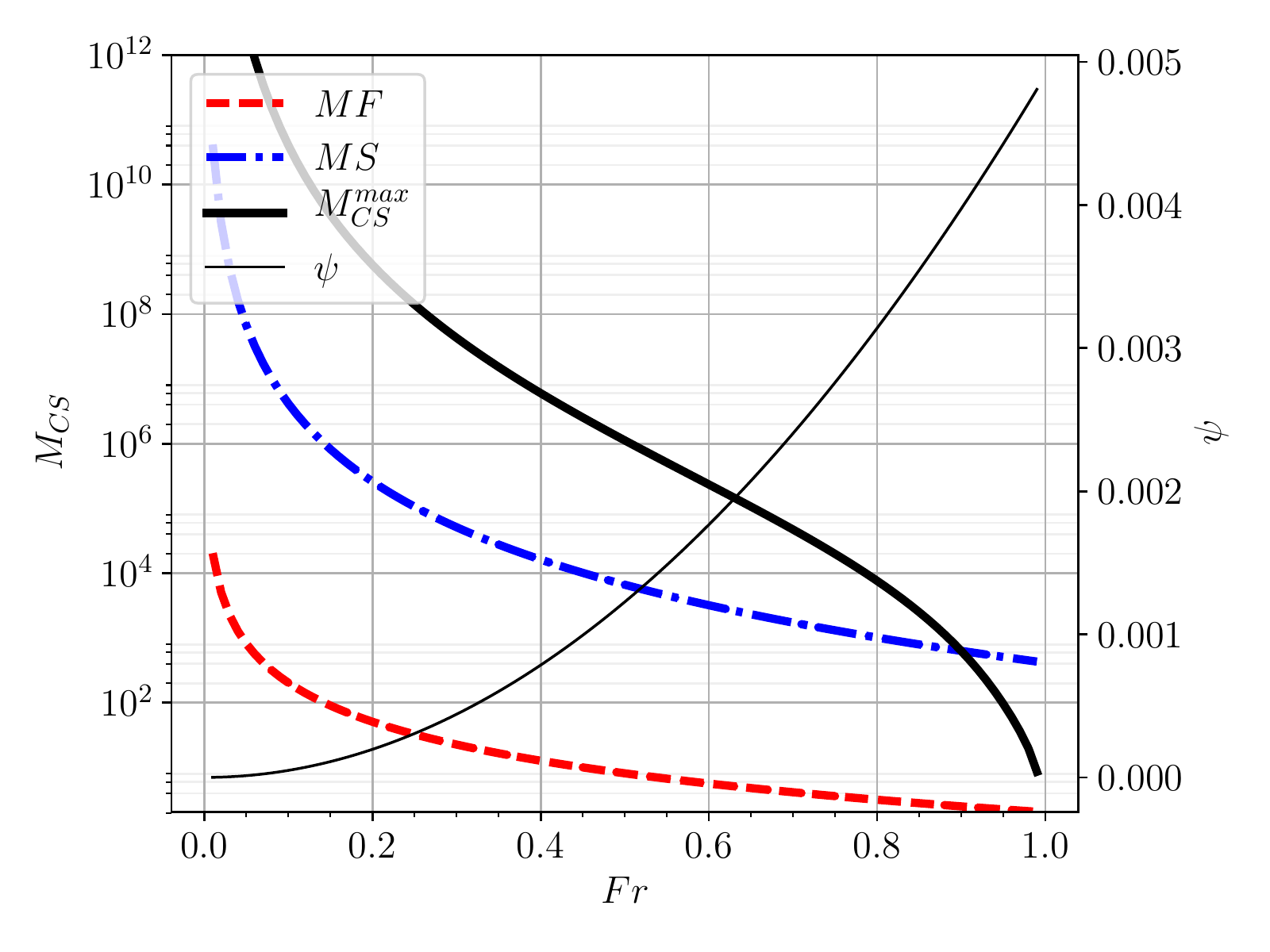}
\includegraphics[scale=0.5]{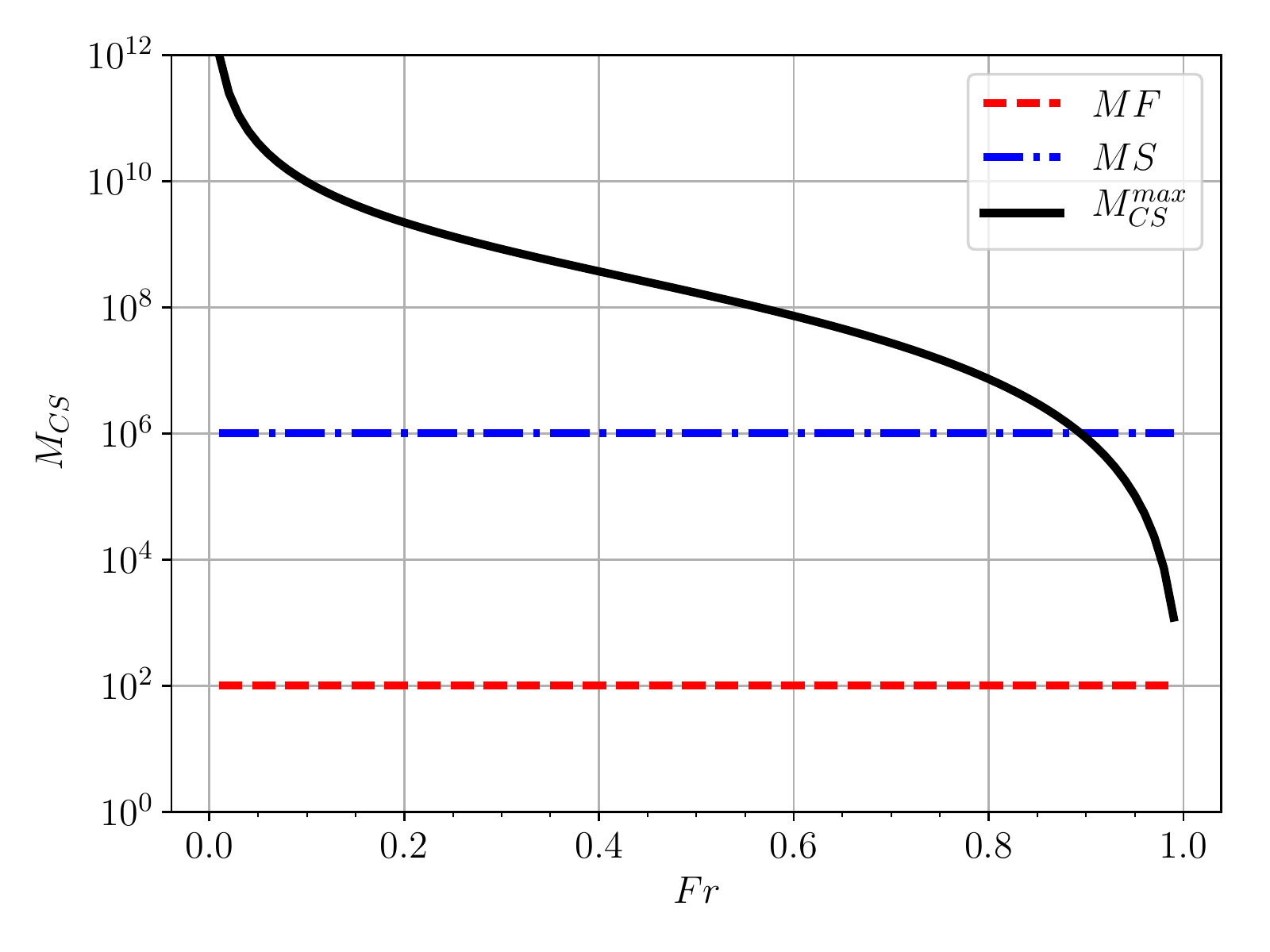}
\caption{Range of validity for the linear morphodynamic acceleration for MORFAC (red dashed line) and MASSPEED (blue dash-dotted line) techniques. In panel a) the curves are obtained by setting $A_g=0.0001$, $\epsilon = 0.01$ in relations (\ref{limit_val_MF}) and (\ref{limit_val_MS}) and using the definition of $\psi$ in (\ref{eq:phi_grass}).
In panel b) the displayed curves are obtained by setting the constant $\psi=10^{-4}$ in relations (\ref{limit_val_MF}) and (\ref{limit_val_MS}).} 
\label{Fig:range_validity}
\end{figure}
\begin{figure}[tb]%
\centering
\includegraphics[width=1\columnwidth]{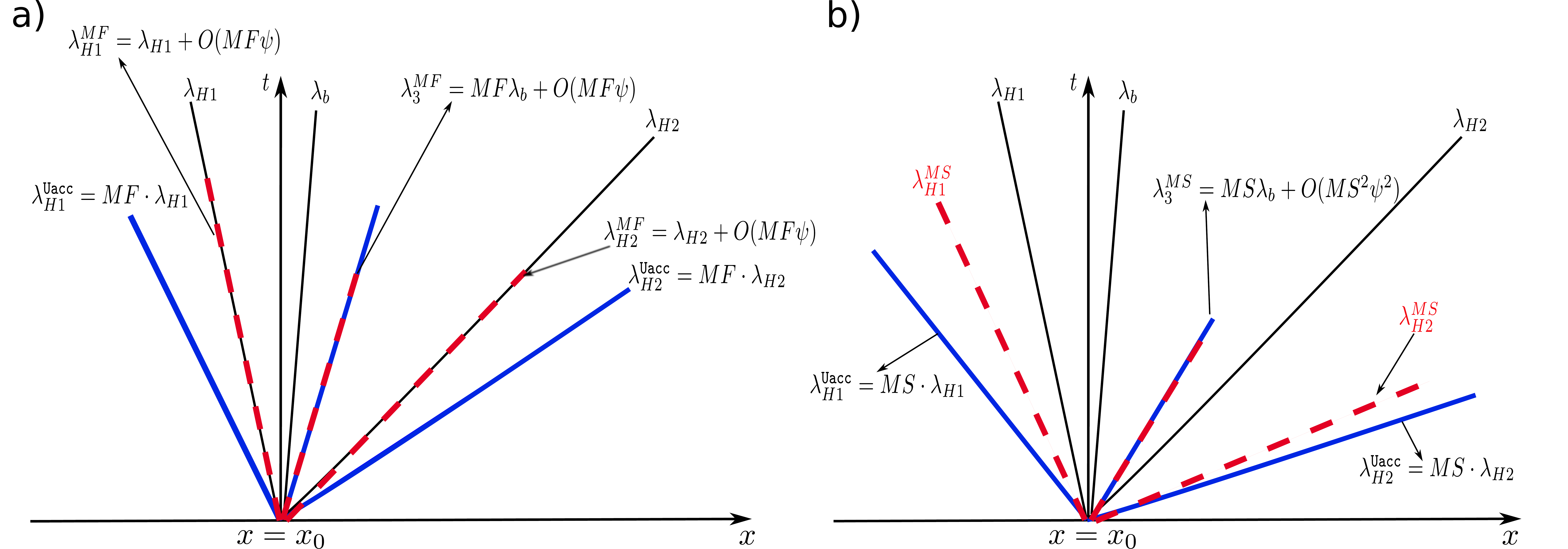}%
\caption{Representation of linearised characteristic curves in the phase space for: original system (black thin lines); uniformly accelerated $\Uacc$ system (blue thick lines); MORFAC approach (red dashed lines, panel a); MASSPEED approach (red dashed lines, panel b). According to the results given in Fig.~\ref{Fig:range_validity} $\MS$ is always larger than $\MF$.}%
\label{fig:MF_vs_MS}%
\end{figure}
%--------------------------------------------
The classical MORFAC approach described in \citet{Roelvink2006} is obtained when the acceleration coefficients are set to $\Mcs=\MF>1$, $\Mq=\Mcw=1$.
Substitution of these values into Eqs.~(\ref{eq_roots_0}) and (\ref{eq_roots_2}) gives 
\begin{equation} \label{eq_root_0_MF}
\frac{\lambda_{1,2}^{\MF}}{c} = \left[ \Fr \mp 1\right]  + O(\MF\,\psi) \, ,
\end{equation} 
\begin{equation} \label{eq_root_2_MF}
\frac{\lambda_3^{\MF}}{c} =   \left[\frac{\Fr}{1-\Fr^2}\right] (\MF\, \psi) -\left[ \frac{\Fr\left(\Fr^2+1\right)}{\left(1-\Fr^2\right)^3}\right] (\MF\, \psi)^2 \;.
\end{equation}  
The accelerated MORFAC system (\ref{eq:NUaccSystem}) can be written in the decoupled form (\ref{eq:def_SW-Ex_decoupled}) if $\lambda_{3}^{\MF}\ll \lambda_{2}^{\MF}$. This condition is satisfied provided that
\begin{equation}
\MF  \, \left[\frac{\Fr}{1-\Fr^2}\right]\, \psi \ll \left [\Fr +1 \right] \;.
\label{limit_val_MF_full}
\end{equation} 
For $\Fr$ numbers typical of environmental flows, i.e. far enough from unity, the term on the left of the inequality (\ref{limit_val_MF_full}) is of $O(\MF\, \psi)$, while the term on the right side is of $O(1)$.
Therefore, the resulting range of validity, for which the decoupled approach holds, is 
\begin{equation}\label{limit_val_MF}
\MF\, \psi = \epsilon
\end{equation} 
where $\epsilon$ is a small parameter ($\epsilon \ll 1$).
If we set $\epsilon=0.01$, use the definition of $\psi(\Fr)$ and set $A_g=0.0001$ s$^2$/m in \eqref{eq:phi_grass}, and plot expression \eqref{limit_val_MF}, we obtain the red dashed line in Fig.~\ref{Fig:range_validity}a. 
This curve identifies the maximum acceleration coefficient $\MF$ that preserves a linear morphodynamic acceleration for a given $\Fr$. Therefore the area below the curve represents the whole range of validity of the decoupled approximation.
It allows that for small $\Fr$ number the acceleration coefficient is very large but rapidly decreases with increasing $\Fr$.
It is also worth noting that, for this particular example, the acceleration factor is smaller than unity for $\Fr \gtrsim 0.6$: this means that the decoupled approximation holds only with a deceleration ($\MF<1$) of the system.
Hence, this represents the value beyond which use of the MORFAC approach becomes counterproductive. 

It is important to underline that the function $\psi(\Fr)$ (thin black line in Fig.~\ref{Fig:range_validity}a) is monotonically increasing but this strongly depends on the chosen transport closure formula (e.g. \ref{eq:phi_grass}).
For the sake of generality we present also the validity range when considering a constant value of $\psi$ (equal to $10^{-4}$ in Fig.~\ref{Fig:range_validity}b).
We, thus, remark that all the subsequent considerations hold both for $\psi(\Fr)$ or $\psi=const$.

If condition (\ref{limit_val_MF}) holds, in agreement with \citet{li2010}, the MORFAC approach does not accelerate hydrodynamics since $\lambda_{1,2}^{\MF}\approx \lambda_{1,2}$ while the bed evolution is linearly accelerated by a factor $\MF$, i.e. $\lambda_{3}^{\MF}= \MF \lambda_{3} + O(\MF \psi)$.
The corresponding situation in the phase space is depicted in Fig.~\ref{fig:MF_vs_MS}a.

Substituting \eqref{eq_root_0_MF} and \eqref{eq_root_2_MF} into definition \eqref{eq:def_Sp}, for a given and constant $\Fr$, the expected computational speed-up of the MORFAC method is
\begin{equation}
\Sp_\MF = \frac{\lambda_3^\MF}{\lambda_3}\cdot \frac{\lambda_{2}}{\lambda_{2}^\MF} \approx \MF \cdot 1  = \MF \;. 
\label{eq:Sp_MF}
\end{equation}

\subsection{The MASSPEED approach}
The MASSPEED ($\MS$) approach is derived from Eq.~\eqref{eq_roots_2}: if $\Mcw=\Mcs=\MS$, the eigenvalue associated to the bed evolution $\lambda_3^\MS$ scales linearly, up to $O(\psi^2)$, with its analogous in the original system \eqref{eq_morpho_root_OPsi}, and the scale factor is $\MS$, thus $\lambda_3^{\MS}=\MS \lambda_3 + O(\psi^2)$.
This suggests that, if the sediment continuity equation (Exner equation) is accelerated by a factor $\Mcs$, an identical acceleration must be imposed on the water continuity equation.
If we set $\Mcw=\Mcs$ in Eqs.~\eqref{eq_roots_2} and \eqref{eq_roots_0} and consider the higher order terms, the condition of weak interaction between bed and hydrodynamics, $\lambda_{3}^{\MS} \ll \lambda_{2}^{\MS}$, is satisfied provided that
\begin{equation}
R  \, \left[\frac{\Fr}{1-\Fr^2}\right]\, \psi \ll \left [\Fr +\sqrt{\Fr^2+R \,(1-\Fr^2)}\right] \quad \text{with}\quad  
R=\frac{\Mcw}{\Mq}=\frac{\Mcs}{\Mq}=\frac{\MS}{\Mq}\,.
\label{limit_val_R_full}
\end{equation} 
For $\Fr$ numbers typical of environmental flows, far enough below unity, the term on the left of the inequality (\ref{limit_val_R_full}) is of order $O(R\psi)$, while the term on the right side is of order $O(R^{1/2})$.
Therefore, the analysis of condition (\ref{limit_val_R_full}) implies it to be satisfied if
\begin{equation}
R \psi^2 = \epsilon\,,\quad \text{with}\quad \epsilon \ll 1 \;.
\label{limit_val_R} 
\end{equation}
This condition can be transformed on a constraint on $\MS$ if a given value is assigned to $\Mq$.
At this stage, we found it convenient to set $\Mq=1$.
Then, we introduce the MASSPEED approach, defined by the following choice for the acceleration coefficients: 
$\Mcw=\Mcs=\MS>1$, $\Mq=1$. Inserting these values into  (\ref{eq_roots_0}) and (\ref{eq_roots_2}) gives 
\begin{equation} \label{eq_root_0_MS}
\frac{\lambda_{1,2}^{\MS}}{c} = \left[\Fr \mp \sqrt{\Fr^2 + \MS \left(1- \Fr^2\right)}\right] + O(\MS\,\psi) \, ,
\end{equation} 
\begin{equation} \label{eq_root_2_MS}
\frac{\lambda_3^{\MS}}{c} = \MS \left[ \left(\frac{\Fr}{1-\Fr^2}\right) \psi -\left( \frac{\Fr\left(\Fr^2+1\right)}{\left(1-\Fr^2\right)^3}\right) \psi^2  \right] \;.
\end{equation}  
According to relation (\ref{limit_val_R}), the condition for the validity of the decoupled approach assumption is satisfied provided that
\begin{equation}\label{limit_val_MS}
\MS \psi^2 = \epsilon\; \text{, with } \epsilon \ll 1 \;.
\end{equation} 
This range is  larger than the analogous range \eqref{limit_val_MF}, valid for the MORFAC approach.
The situation is displayed in Fig.~\ref{Fig:range_validity}, where the  blue dash-dotted line is obtained by setting $\epsilon =0.01$ in both panels.
The results is that for all the range of $\Fr$, the MASSPEED allows for larger values of the acceleration coefficient than those given by the MORFAC approach.
The areas below these lines represent the range of validity of the decoupled approximation.

Different from the MORFAC approach, the MASSPEED acceleration modifies also the characteristics of hydrodynamics, i.e. $|\lambda_{1,2}^{\MS}| > |\lambda_{1,2}|$ (compare Figures \ref{fig:MF_vs_MS}a and \ref{fig:MF_vs_MS}b).
In particular, according to \eqref{eq:def_Sp}, $\mathtt{R_H} = \lambda_{2}^\MS/\lambda_{2} > 1$ hence the theoretical speed-up is bounded as follows:
\begin{equation}\label{bounds_MS}
1<\Sp_\MS = \frac{\mathtt{R_M}}{\mathtt{R_H}} \approx \frac{\MS}{\mathtt{R_H}}  < \MS.
\end{equation}
 
This may wrongly suggest that the MORFAC gives a larger speed-up than MASSPEED, while it is true that the MASSPEED approach allows for much larger values of $\Mcs$ which compensates by far the reduction of the integration time-step due to the acceleration of the hydrodynamic characteristic $\lambda_{2}^\MS$. 

Since $\lambda_3^{\MS}$ is increased by the rate of $\mathtt{R_M}$ and $\lambda_2^{\MS}$ by the rate $\mathtt{R_H}<\mathtt{R_M}$, there is the risk that small morphodynamic disturbances may be accelerated so as to travel faster than the fastest hydrodynamic small disturbances, i.e. $\lambda_3^{\MS}\ge\lambda_2^{\MS}$.    
Therefore it sounds reasonable to impose a physical limit for the validity of the MASSPEED approach. 
The physical limit of such an acceleration is given by the fact that the bed wave perturbation associated to $\lambda_3^{\MS}$ should travel at a slower pace than  the perturbation of the water surface  associated to $\lambda_2^{\MS}$. 
Now, if we consider the approximation of $\lambda_2^{\MS}$, Eq.~\eqref{eq_root_0_MS}, at the leading order $O(\psi^0)$ and $\lambda_3^{\MS}$, Eq.~\eqref{eq_root_2_MS}, at $O (\psi)$, and impose $\lambda_2^{\MS} = \lambda_3^{\MS}$, we obtain the following limit relation
\begin{equation} \label{Mcs_max}
	\MS^{\max}= -\frac{\left(\Fr^6+2 \Fr^4 \psi-3 \Fr^4-2\Fr^2 \psi+3 \Fr^2-1\right)}{\Fr^2 \psi^2},
\end{equation}
which gives the maximum factor $\MS^{\max}$ that can be used to avoid this unphysical behaviour.
Accelerating the system by a factor $\MS<\MS^{\max} $ avoids the loss of the strict hyperbolicity, which occurs when two eigenvalues coalesce \cite{Cordier2011,Toro2009}.
This particular mathematical condition must be avoided because may give rise to resonance and loss of solution uniqueness.
Theoretical issues regarding resonance are found in the classical papers \cite{isaacson1992,liu1987} and references therein.
In Fig.~\ref{Fig:range_validity} the condition of the physical validity of \eqref{Mcs_max} is plotted as a function of the Froude number.
It is seen that for $\epsilon=0.01$ the range of validity for the decoupled solution is contained within the limit of physical validity for all $\Fr$ values. 

\begin{figure}[ht]
\centering
\includegraphics[scale=0.6]{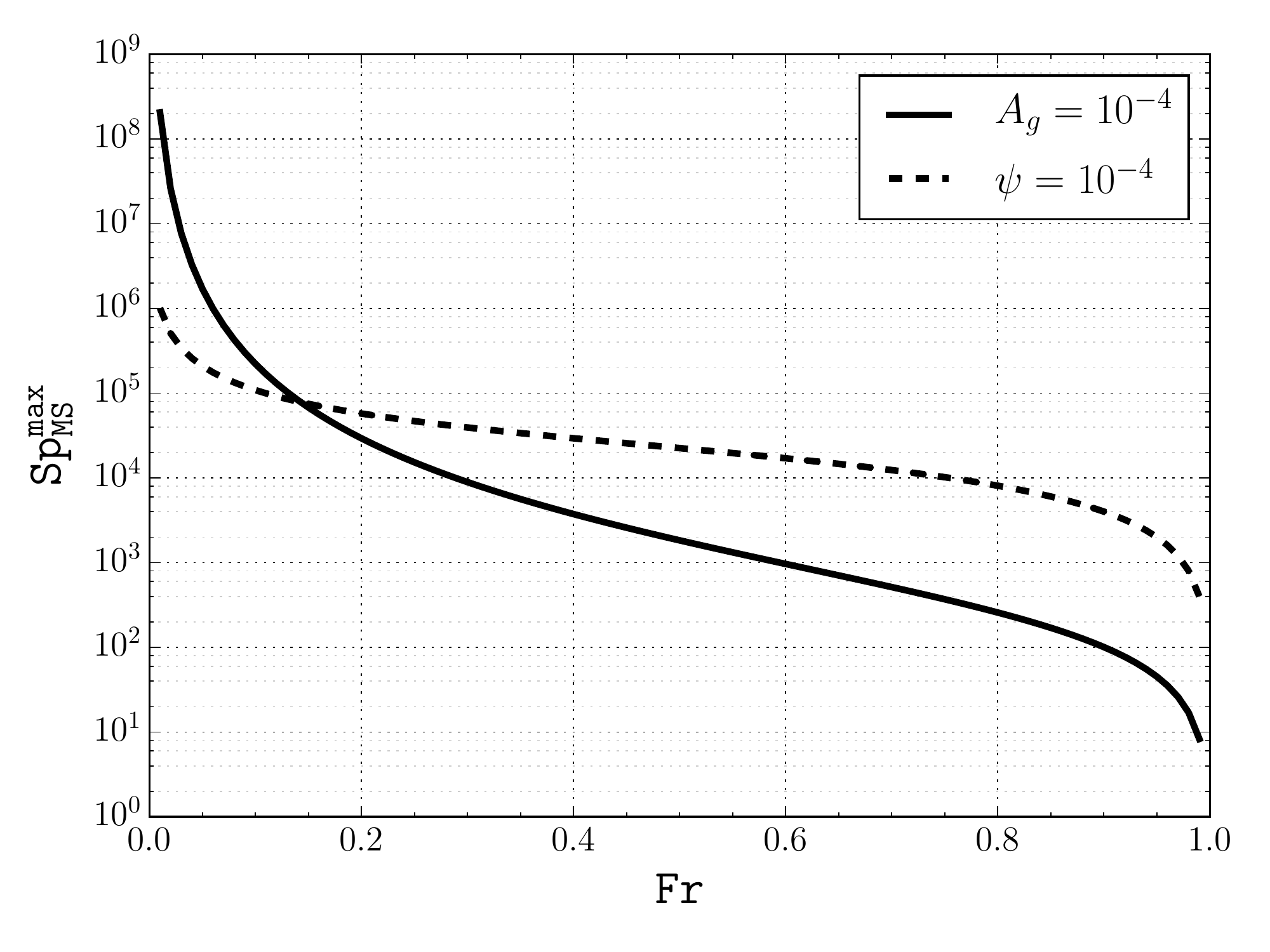}
\caption{MASSSPEED approach: maximum theoretical speed-up $\Sp_\MS^{\max}$ (Eq.~\ref{Gain_lin_max}) obtained from the approximate solution of the eigenvalues. 
The full line is obtained by using the definition of $\psi$ as in Eq.~(\ref{eq:phi_grass}) and setting $A_g=0.0001$ in the sediment transport closure (\ref{eq:Grass}). The dotted line is obtained by setting $\psi=10^{-4}$.}
\label{Fig:Max_gain}
\end{figure}

Finally, substitution of Eq.~\eqref{Mcs_max} in the definition of the maximum speed-up gives
\begin{equation} \label{Gain_lin_max}
\Sp^{\max}_\MS=
	\MS^{\max}\frac{1+\dfrac{1}{\Fr}}{1+\sqrt{1+\MS^{\max}\left(\dfrac{1}{\Fr^2}-1\right)}}.
\end{equation}
If we substitute the definition of $\psi$ and set $A_g=0.0001$ in \eqref{eq:Grass}, and plot relation \eqref{Gain_lin_max}, we obtain the black solid curve in Fig.~\ref{Fig:Max_gain}.
Moreover, the dashed line is obtained by setting the constant $\psi=10^{-4}$ in \eqref{Gain_lin_max}.  
It is observed that the speed-up is of order $10^2$ for small Froude numbers and decreases rapidly to unity as $\Fr\rightarrow1$.

\section{Non-linear numerical strategies to compute the maximum acceleration factors}
\label{sec:Non_Lin_An}
In the previous section the maximum accelerations allowed with MORFAC and MASSPEED approaches have been introduced thanks to a linear approximation of the system eigenvalues (see \S \ref{sec:approx_sol_NU}).
The goal of this section is to defined a more general criterion for the determination of the  maximum acceleration factors by considering the fully nonlinear expression of the three eigenvalues as described in \ref{apx_A}.
Thus, we propose a further strategy to dynamically recompute  the maximum acceleration factors during a numerical simulation. If not differently specified, we consider an acceleration factor $\Mcs=10$ and $A_g=0.005$ in \eqref{eq:Grass}.  

\subsection{Non-linear estimation of the maximum acceleration factors}
\label{sub:OptimalMORFAC}
\begin{figure}[p]%
	\centering
	\subfloat 
	{\includegraphics[width=0.48\columnwidth]{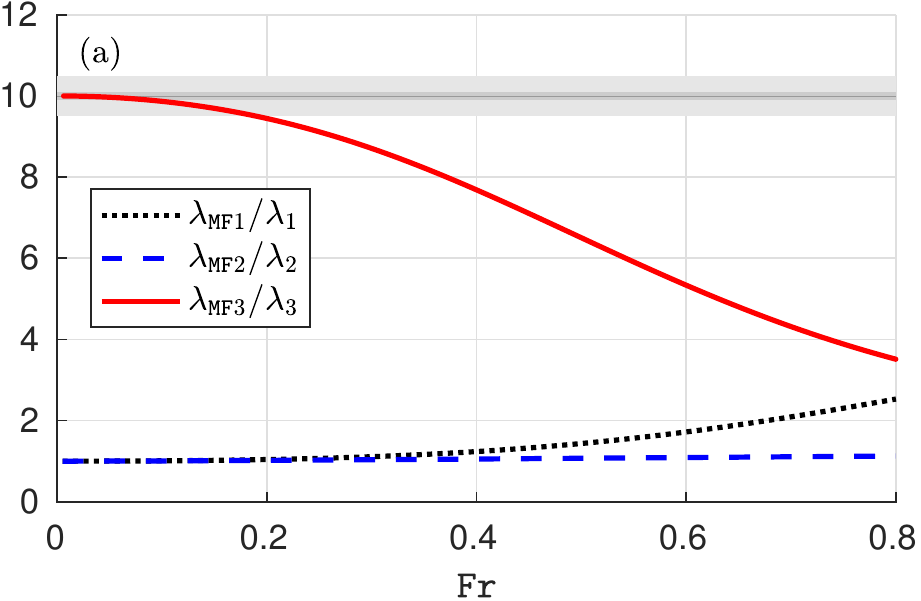}}\quad
	\subfloat 
	{\includegraphics[width=0.48\columnwidth]{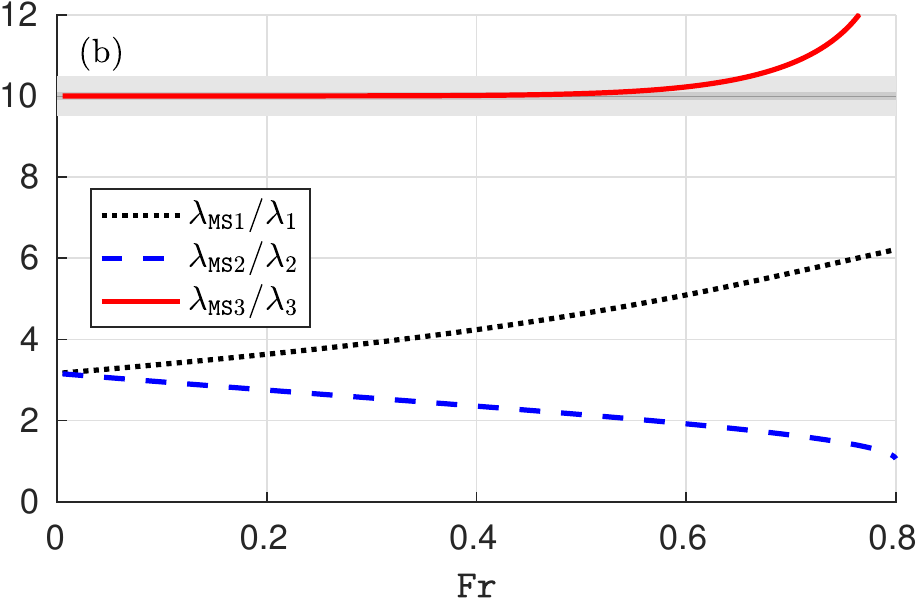}}\\%
	\subfloat 
	{\includegraphics[width=0.48\columnwidth]{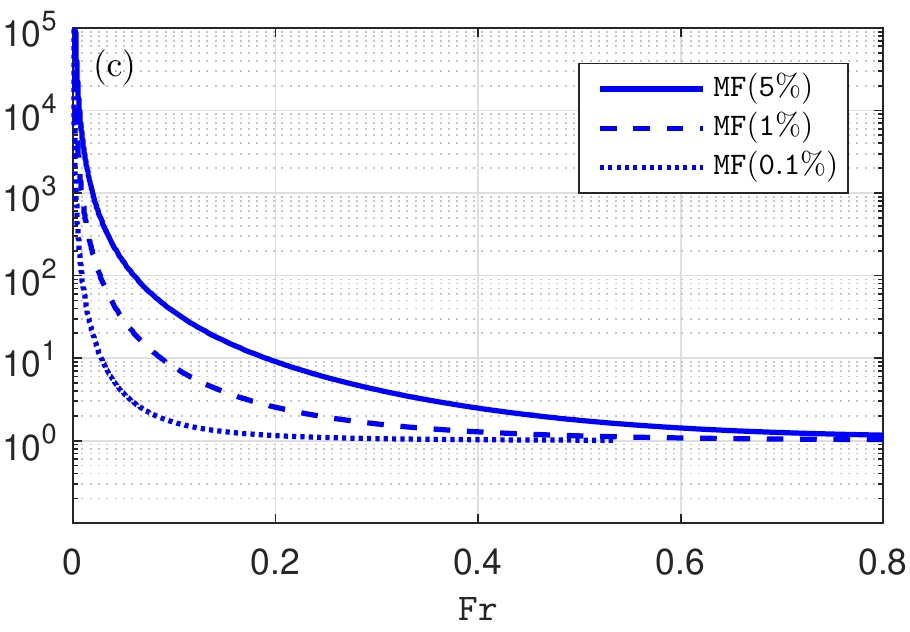}}\quad
	\subfloat 
	{\includegraphics[width=0.48\columnwidth]{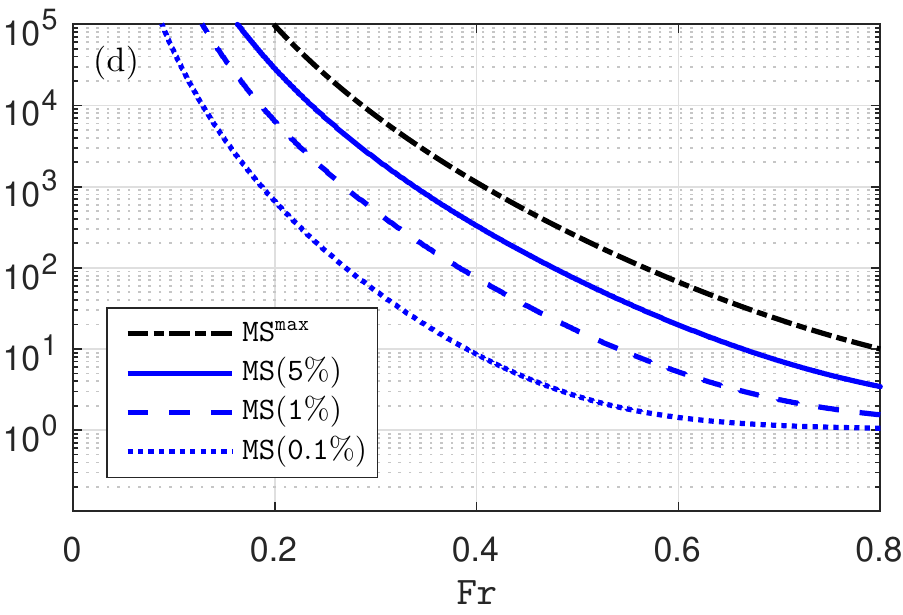}}\\%
	\subfloat 
	{\includegraphics[width=0.48\columnwidth]{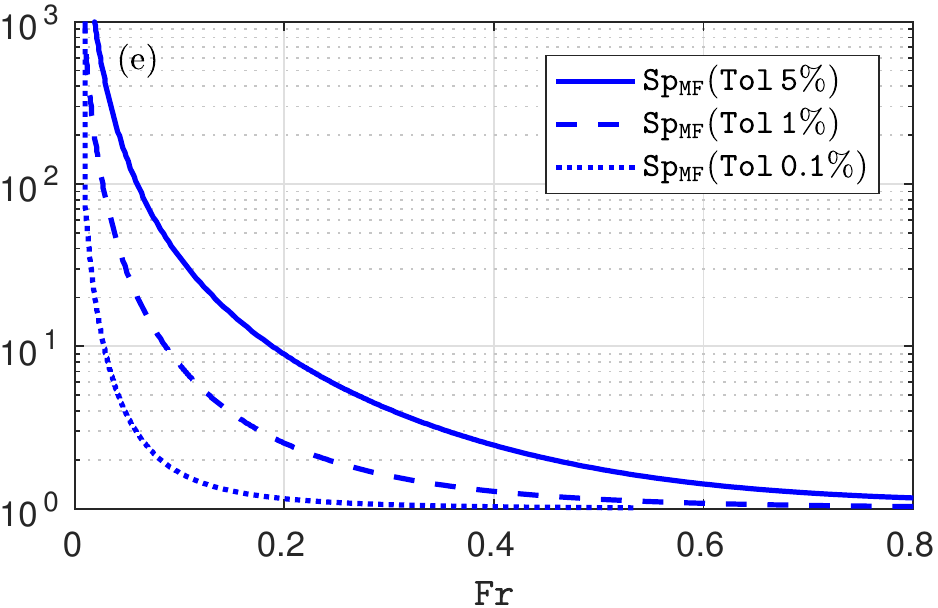}}\quad
	\subfloat 
	{\includegraphics[width=0.48\columnwidth]{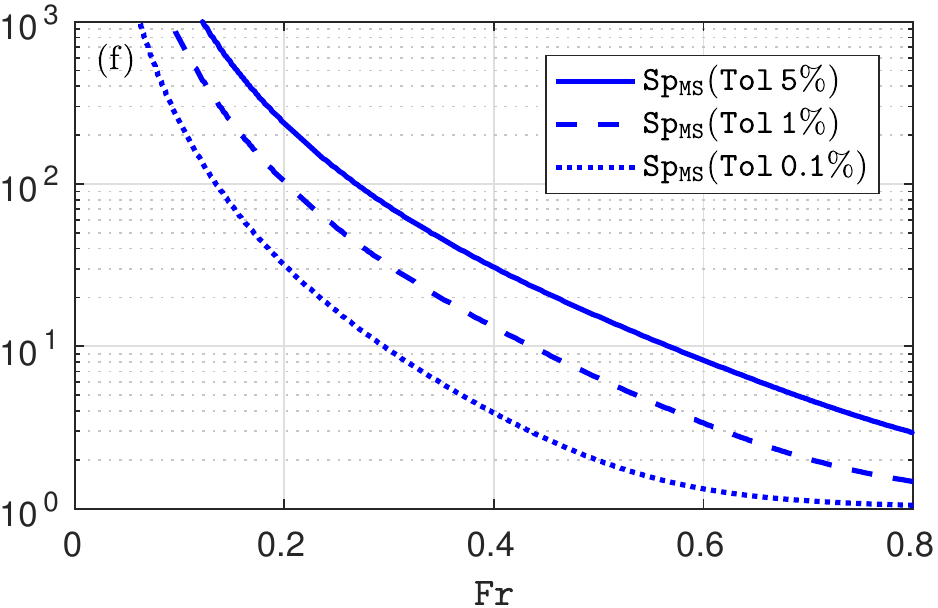}}\\%
	\caption[]{Panel a and b: Ratios of nonlinear eigenvalues $\frac{\lambda_{\MF i}(\Fr,\MF,\psi)}{\lambda_i(\Fr,\psi)}$. The grey-shaded areas represents the tolerance band $\Tol$ in Eq.~\eqref{eq:linexp}. 
Panels c and d: maximum acceleration factor  for given values of tolerance.   Panels e and f: maximum expected speed-up. In the left and right panels the results for the MORFAC and MASSPEED approaches are collected.}%
	\label{fig:MORSvsMORF}%
\end{figure}
%----------------------------------------------------------------------------

First we consider the MORFAC approach.
The roots associated with the flux matrix $\M\,\A$ in Eq.~\eqref{eq:NUaccSystem}, can be computed by solving the cubic characteristic polynomial  \eqref{eq_char_pol_Uacc} after setting the acceleration factors equal to $\Mcw=\Mq=1$ and $\Mcs=\MF = 10$, in this example.
The MORFAC approach successfully establishes a well defined correspondence between the original and the accelerated model only if morphodynamics evolves linearly in time.
More precisely, under the hypothesis of weak interaction between hydrodynamics and morphodynamics, the following relation must hold:
\begin{equation}\label{eq:linapr}
\mathtt{R_M} = \frac{\lambda_{\MF 3}(\Fr,\MF,\psi)}{\lambda_3(\Fr,\psi)} \approx \MF,
\end{equation}
where $\lambda_3$ is the smallest eigenvalue (the morphodynamic eigenvalue) of the original system.
Hence, the more $\mathtt{R_M}$  deviates from the assigned $\MF$ , the weaker the assumption of linear acceleration is.
We quantify the maximum acceptable deviation introducing a tolerance band ($\pm\Tol$), therefore condition \eqref{eq:linapr} can be rewritten as
\begin{equation}\label{eq:linexp}
\mathtt{R_M} =\frac{\lambda_{\MF 3}(\Fr,\MF,\psi)}{\lambda_3(\Fr,\psi)} = \MF\, (1 \pm \Tol) \;.
\end{equation}
Relation \eqref{eq:linexp} is an implicit expression of $\MF$ that depends on the water flow ($\Fr$) and sediment transport ($\psi$).
Hence, for a user-given tolerance ($\Tol$), the maximum value of $\MF$, which assures that the bed evolves linearly, is numerically computed from Eq.~\eqref{eq:linexp}. 
In Figure \ref{fig:MORSvsMORF}a, the ratios $\frac{\lambda_{\MF i}(\Fr,\MF,\psi)}{\lambda_i(\Fr,\psi)}$ are displayed by using different style-lines.
The ratio $\lambda_{\MF3}/\lambda_3$ (red solid line), tends asymptotically to the assigned $\MF$ as $\Fr\rightarrow0$.
On the other hand, for increasing values of $\Fr$ the ratio rapidly decays below $\MF=10$.
The linear assumption  holds as long as the red curve lays inside the grey-shaded areas, corresponding to the right-hand-side of \eqref{eq:linexp}: the thinner and the darker the gray stripe is the smaller is the tolerance, $\Tol= 5\%,\,1\%,\,0.1\%$ in this case. 

In Fig.~\ref{fig:MORSvsMORF}c the maximum acceleration factor $\MF$, computed  with \eqref{eq:linexp}, is plotted against  $\Fr$. The three different lines are obtained by considering three different small tolerance values, namely $\Tol= 5\%,\,1\%,\,0.1\%$.
The acceleration coefficient $\MF$ shows an inverse exponential dependency on $\Fr$.
This is in agreement with the empirical results obtained by \citet{Ranasinghe2011} and \citet{li2010}.
It is also seen that the magnitude of the maximum acceleration factor crucially depends on the user-given tolerance. 

Finally, for each value $\MF$, the expected numerical speed-up $\Sp_\MF$ \eqref{eq:def_Sp} is plotted against $\Fr$ in Fig.~\ref{fig:MORSvsMORF}e. As expected, the speed-up is decreasing rapidly for increasing values of $\Fr$. For example, if we consider $\Fr=0.1$ and  set the tolerance to $1\%$, the maximum achievable speed-up is about 10 while if we increase the tolerance up to $5\%$ the corresponding speed-up increases up to about 100.  

The very same analysis can be extended to the MASSPEED approach, providing $\Mq=1$ and $\Mcw=\Mcs=\MS$ in the governing system \eqref{eq:NUaccSystem}.
The maximum MASSPEED factor $\MS$ for given $\Fr$, $\psi$ and tolerance $\Tol$ is analogous to \eqref{eq:linexp} and reads
\begin{equation}\label{eq:linexpMS}
\mathtt{R_M} = \frac{\lambda_{\MS 3}(\Fr,\MS, \psi)}{\lambda_3(\Fr,\psi)} = \MS\, (1 \pm \Tol) \;.			
\end{equation}
It is interesting to note that, consistently with the results obtained with approximated solutions in \S \ref{sec:LmaT}, the range of $\Fr$ numbers for which the ratio $\lambda_{\MS 3}/{\lambda_3}\simeq \MS$ (range of linearity), is broader for the MASSPEED when compared with the MORFAC approach.
Comparing Fig.~\ref{fig:MORSvsMORF}a and Fig.~\ref{fig:MORSvsMORF}b, the linear range extends up to  $\Fr\approx 0.6$ (panel b) for the MORSPEED, while it reduces to $\Fr\approx0.15$ for the MORFAC (panel a).
Within the linear range, the MASSPEED approach shows also higher values of the maximum acceleration (Fig.~\ref{fig:MORSvsMORF}c versus \ref{fig:MORSvsMORF}d) and larger speed-up  (Fig.~\ref{fig:MORSvsMORF}e versus \ref{fig:MORSvsMORF}f). 
For example, given a tolerance of $1\%$ and $\Fr=0.4$, the maximum $\MS$ corresponds to 75 (Fig.~\ref{fig:MORSvsMORF}d) resulting in a speed-up of about 13 (Fig.~\ref{fig:MORSvsMORF}e) while application of the MORFAC approach does not result in any acceleration. 

Finally, concerning the loss of hyperbolicity, we compute the value $\MS^{\max}$ by imposing that $\lambda_{\MS2} = \lambda_{\MS3}$ and making use of the fully nonlinear expression of the eigenvalues. $\MS^{\max}$ is plotted against the $\Fr$ number in Fig.~\ref{fig:MORSvsMORF}d for a given tolerance of $5\%$ (black dashed line). It is worth noting that the loss of hyperbolicity occurs outside the domain of linear acceleration, i.e. $\MS^{\max}>\MS(5\%)$.

\subsection{Numerical evaluation of the highest acceleration factor: fixed and adaptive appraoch}\label{fix_adaptive_sec}
Here we propose a numerical strategy similar to the well-known Courant-Friedrichs-Lewy stability condition to maximize the computational speed-up. Note that here we refer only to MASSPEED approach, given that the very same procedure can be implemented for the MORFAC approach.

Let us consider a physical domain of length $L$ discretized with  a finite  set  of points or volumes, regardless if finite difference or finite volume approaches are used. 
At a given time, within a single numerical time step $\Delta t$, local flow ($\Fr_i$) and sediment transport ($\psi_i$) conditions are assigned and therefore the calculation for each cell $i$ of the nonlinear eigenvalues $\lambda_{\MS j,i}$ and $\lambda_{j,i}$ (with $j=1,2,3$) is possible. 
Then, for a given tolerance value $\Tol$ (prescribed by the user), application of relation (\ref{eq:linexpMS}) gives the value of the maximum accelerator factor $\MS_i$ for each cell $i$ at a given time. 
Now, two possible approaches are introduced here: fixed and adaptive.
In the fixed approach, the maximum acceleration factor $\MS$ is computed at the beginning of the simulation as $\MS=\min\limits_i{\left[  \MS_i \right]  }$ and kept constant for all time steps of the simulation, until the final time is reached.
In the adaptive approach the maximum acceleration factor is a function of time, i.e. $\MS_i(\tau)$, and is computed for each time step according to the local flow and sediment transport conditions. 
Then, for the generic time $\tau$, the adaptive $\MS(\tau)=\min\limits_i{\left[ \MS_i(\tau) \right]}$ is computed solving the following equation for each cell $i$: 
\begin{equation}
		\frac{\lambda_{\MS 3,i}\big(\Fr_i(\tau),\psi_i(\tau),\MS_i(\tau)\big)}{\lambda_{3,i}\big(\Fr_i(\tau),\psi_i(\tau)\big)} = (1\pm\Tol)\cdot\MS_i(\tau)\,.
		\label{eq:Adaptive-CFL1}
\end{equation}
Eq.~\eqref{eq:Adaptive-CFL1} is valid for any closure for the solid transport and can be solved via a numeric iterative method (e.g. standard regula falsi (RF) method \cite{iterativeMethods}). 
Alternatively, to reduce the computational cost due to the iterative procedure, one can obtain $\MS_i(\tau)$ followed for obtaining the curves in Fig.~\ref{fig:MF_vs_MS}.

\section{Evolution of a sediment hump: linear and numerical solutions}
\label{sec:application}
We assess and compare the accuracy and efficiency of the MORFAC and the new  linear morphodynamic accelerator MASSPEED by considering the  propagation of a sediment hump.
In all cases, the solutions obtained with MORFAC and MASSPEED are compared with that obtained with the original model.
First, we solve a linearized problem for which an analytical solution is available. We use this problem as a \textit{proof of concepts} of the theoretical backgound we developed.  
Second, we solve numerically the long-term evolution of a sediment hump.
The final goal is to assess the advantages of the new MASSPEED approach.
Numerical integration is performed by using a classical one-dimensional finite volume scheme.
We adopt a path-conservative solver of the DOT \cite{Dumbser2011} type where the use of the analytical formulation of the eigenstructure of the system flux matrix improves the computational efficiency \cite{Carraro2016,ADOT}. 

\subsection{Linearized morphodynamic problem: evolution of a small sediment hump}
\label{sec:MathSol}
%-------------------------------------------------------------
\begin{figure}[t]%
\centering
\includegraphics[width=\columnwidth]{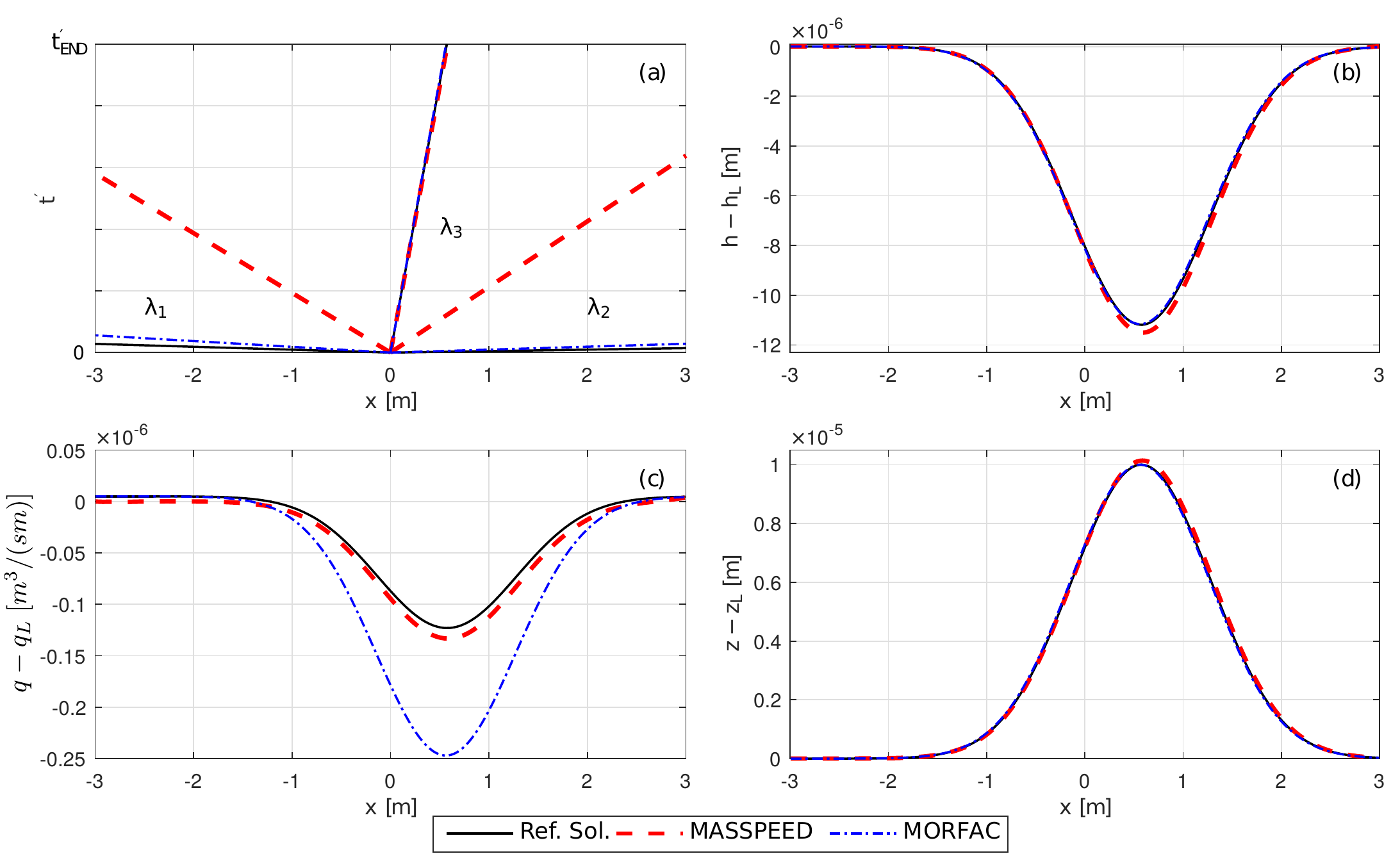}
\caption{Linearized solution of the propagation of a sediment hump \cite{Lyn2002} using the linearized MORFAC and MASSPEED approaches: a) comparison between the characteristic lines (the time is properly scaled for the MORFAC and MASSPEED approaches); b) difference between the water depth and the unperturbed water depth; c) difference between the flow discharge and the unperturbed flow discharge; d) bottom elevation. The analytical solution of the linear original system is used as reference.}
\label{fig:LinearSol}%
\end{figure}
%-------------------------------------------------------------
We consider the evolution of a small  (infinitesimal) erodible hump due to a nearly uniform water flow in a straight channel. Under these conditions, the problem can be studied within a linear framework, and  an analytical solution can be easily derived \cite[e.g.][]{Lyn2002}. We consider the original nonlinear system \eqref{eq:NUaccSystem} (the friction  term is neglected) and perform a linearization by \textit{freezing} the flux matrix \eqref{eq:Jacobian}, or more specifically its entries, considering the following uniform reference state, $\W_\U=[h_\U,\,q_\U,\, z_\U]^T$. The resulting  \emph{linearized system} is
\begin{equation}
\frac{\partial \W}{\partial t}+\M\,\A_\U\frac{\partial\W}{\partial x} = 0\,,
	\label{eq:LinearSystem}
\end{equation}
with
\begin{equation}
	\A_\U=
	\begin{bmatrix}
		0                   &	1	        &	0       \\
		c_\U^2-u_\U^2	        &   2u_\U    &	c_\U^2		\\
		-u_\U\,\psi_\U 		&	\psi_\U		&	0
	\end{bmatrix}	\qquad \qquad
	 \M = \begin{bmatrix}
					\Mcw	&	0	& 0 		\\
					0	&	1		&	0		\\
					0	&	0  		&	\Mcs
			\end{bmatrix} \;.
	\label{eq:linear_A}
\end{equation}
%$\mathtt{Fr}_\U$
The original system is obtained when $\Mcw=\Mcs=1$, the MORFAC system for $\Mcw=1$, $\Mcs=\MF$ and the MASSPEED for $\Mcw=\Mcs=\MS$. 
Subscript $\U$ refers to the reference state and $u_\U=q_\U/h_\U$ and $c_\U=\sqrt{g\,h_\U}$ are the reference flow velocity and celerity, respectively. $\psi_\U=\psi|_{\W_\U}=3\, g\, \xi\, A_g\, \mathtt{Fr}_\U^2$ is the reference sediment transport parameter where $\Fr_\U = u_\U/c_\U$ is the reference Froude number and $q_0 = q_\U = \sqrt{h_\U^3\,g\,\mathtt{Fr}_\U^2}$. The initial conditions of the problem are given by $\W_0=[h_0,\,q_0,\, z_0]^T$ with $z_0(x) = z_{\max} \exp(-x^2)$, $z_{\max}=1.0^{-5}$ m and $h_0(x)=h_\U - z_0(x)$. We also assume $\psi_\U = 0.01$ and  $h_\U=1$ m to which correspond $\Fr_\U = 0.33$.
The solution of system \eqref{eq:LinearSystem} can be obtained analytically by using characteristic variables \cite{Toro2009} as done by \citet{Lyn2002}. More details on how the linear analytical solution is obtained are given in \ref{app:Val}. 
The final time for which the solution is sought is $t_\mathtt{END}=50$~s for the original system.
Since the infinitesimal amplitude of the hump, linear conditions hold and therefore the time scales of the accelerated  and original system are linked by a linear relation, i.e.~the corresponding final time of the simulation for the MORFAC and MASSPEED approaches yields $t_\mathtt{END,s}=t_\mathtt{END}/\MF$ and $t_\mathtt{END,s}=t_\mathtt{END}/\MS$, respectively.   
Setting a tolerance $\Tol=1.36\%$, the resulting (rounded) maximum acceleration factors, calculated from Eqs.~\eqref{eq:linexp} and \eqref{eq:linexpMS}, are $\MF=2$ and $\MS=900$.
The analytical solutions are displayed in Fig.~\ref{fig:LinearSol}.  In  Fig.~\ref{fig:LinearSol}a the characteristic lines in the phase space $x-t'$ are given, where $t'$ stands for $t$ the time in the original system, $t/\MF$ and $t/\MS$, the times in the MORFAC and MASSPEED systems.  Deviations of the water depth, discharge and bed elevation, to their unperturbed initial values $h_L$, $q_L$ and $z_L$ are given in Fig.~\ref{fig:LinearSol}b, Fig.~\ref{fig:LinearSol}c and Fig.~\ref{fig:LinearSol}d, respectively. 

It is seen in the phase space plots, the three eigenvalues with the propagation of bed information, $\lambda_3\equiv\lambda^\MF_3\equiv\lambda^\MS_3$, overlap each other (Fig.~\ref{fig:LinearSol}a).
As a consequence the celerity of the infinitesimal bed form is well predicted by both the MORFAC and MASSPEED approaches (see also Fig.~\ref{fig:R32comp} in \ref{app:Val}).
It is also seen that the amplitude of water depth (Fig.~\ref{fig:LinearSol}b) and bed elevation (Fig.~\ref{fig:LinearSol}c) are also accurately predicted.
A marked difference is found for the water discharge (Fig.~\ref{fig:LinearSol}c) for which the MORFAC approach introduces a larger deviation to that of MASSPEED. 

\subsection{Long term evolution of a bottom hump: frictionless case}
\label{sub:hump}
%---------------------------------------------------------------------%
\begin{figure}[tb]%
\centering
\includegraphics[width=0.5\columnwidth]{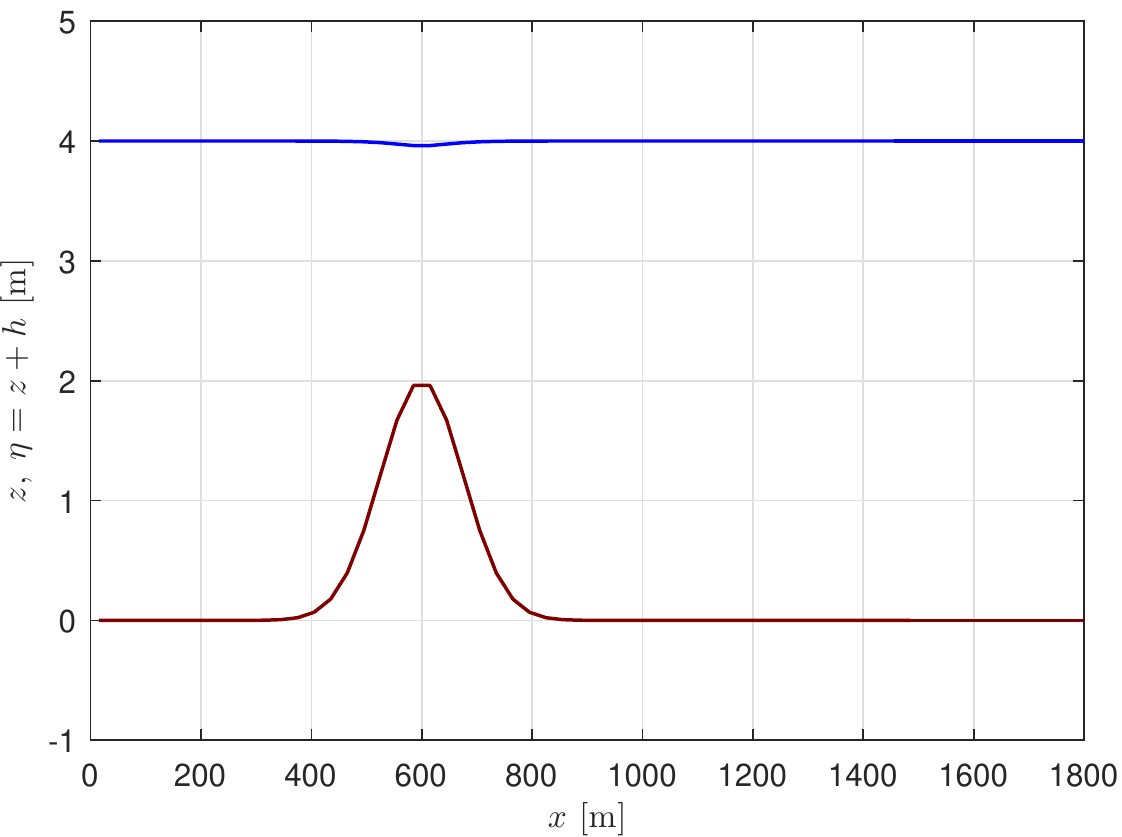}%
\caption{Long term evolution of a bottom hump: initial water surface elevation (blue) and bed elevation (red) profiles.}%
\label{fig:HumpIC}%
\end{figure}
%---------------------------------------------------------------------%
\begin{figure}[tb]%
	\centering
	\subfloat 
	{\includegraphics[width=0.48\columnwidth]{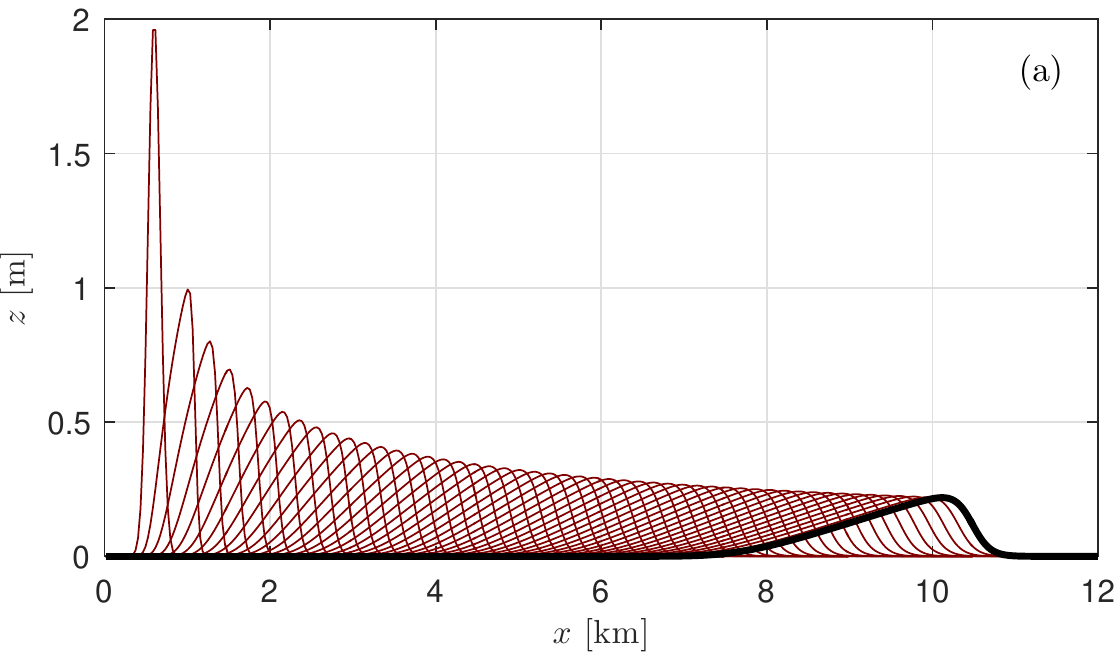}}\quad
	\subfloat 
	{\includegraphics[width=0.485\columnwidth]{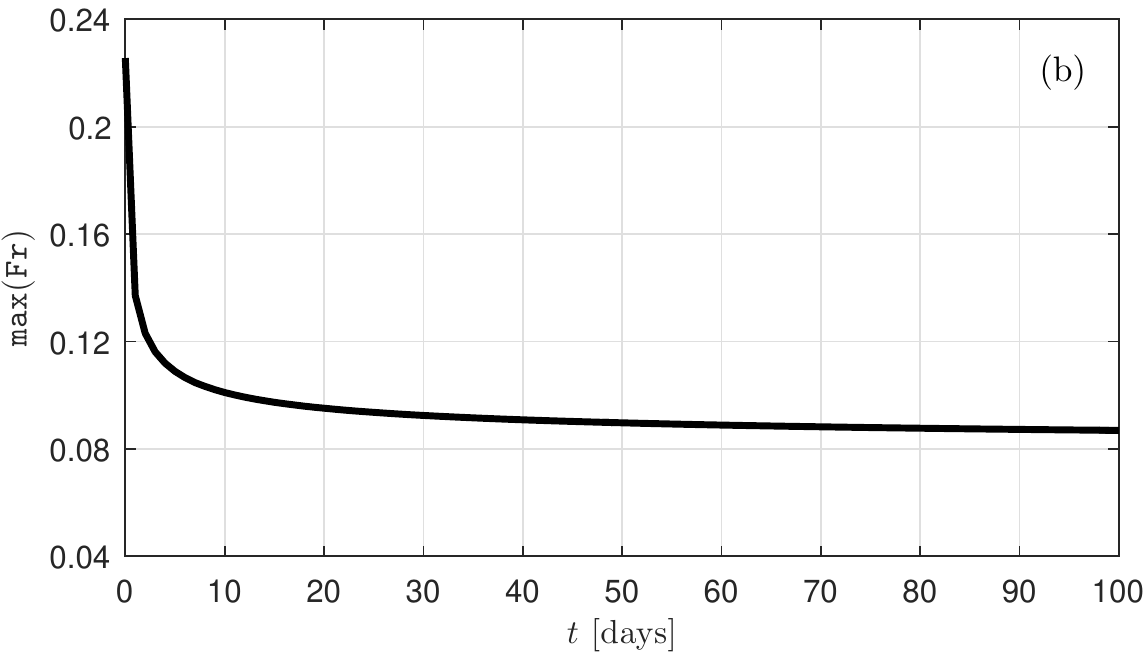}}\\
\caption{Reference solution for the long term evolution of a bottom hump: a) bottom profiles every two days, in thick line  day number 100; b) time evolution of the maximum $\Fr$ measured within the numerical domain.}%
\label{fig:BedFrEvo}%
\end{figure}
%---------------------------------------------------------------------%
This test case consists of the simulation of the long term evolution of an erodible bottom hump immersed into a quasi-steady, frictionless flow; it is similar to the test proposed by \citet{Ranasinghe2011}. 
The initial bed profile is given by 
\begin{equation}
	z(x,0)= z_{\max}\, \exp\left[{-\frac{(x-\mu)^2}{\sigma^2}}\right],
	\label{eq:HumpIC}
\end{equation}
where $z_{\max} = 2$ m, $\mu = 600$ m and $\sigma = 150$ m.
The initial uniform water discharge is $q(x,0)=q_0=2$ m$^3$/(s m) while the initial water depth corresponds to the steady state water profile for the given discharge and initial bed topography.
A constant discharge $q_0$ is imposed at the upstream boundary while a constant water depth $h_0=4$ m is imposed at the downstream boundary.
Fig.~\ref{fig:HumpIC} shows the initial flow and bed configuration.
Transmissive downstream boundary conditions are imposed for the bed and $A_g$ is set  to 0.005 s$^2$/m in the sediment transport formula \eqref{eq:Grass}.
The numerical domain ($x \in [0;12000]$ m) is discretized with 400 cells of constant width ($30$ m).
Finally, ${t}$ the morphodynamic output time is set equal to 100 days.

Fig.~\ref{fig:BedFrEvo}a shows the bottom evolution in time, with a temporal breakdown of 2 days, obtained with the original, non-accelerated model.
In the first simulated days, the amplitude of the hump decreases rapidly and after about 20 days the bottom assumes a flatter and stable profile.  
The decrease of the hump amplitude is associated to a corresponding decrease of the maximum Froude number on the computational domain (Fig.~\ref{fig:BedFrEvo}b).

\begin{table}[tb]%
\begin{center}
\caption{Long term evolution of the bottom hump: acceleration parameters ($\MF$, $\MS$ and $\Tol$), number of computational time steps, CPU time, numerical \eqref{eq:def_numSp} and theoretical \eqref{eq:def_Sp} speed-up, normalized error $\Eg$ and final position of the hump crest.} 
\begin{tabular}{lllrrcccc}
\toprule
{\#} &	{Method}	&	{Acc. Factors}	&	{Time steps}	&	{CPU [s]}	&	{$\Sp^\mathtt{CPU}$}	&	{$\Sp$}	&	{$\Eg$}	&	{x(crest) [m]}	\\
\midrule
{0}& {Ref. Sol.}		&				{}				              &		2 050 300	&		   6803	&	 {-}	&	 {-}		&	{-}		&	10125	\\
\midrule
{1}&\multirow{ 3}{*}{MORFAC}			&	{$\MF_{5.0\%}=7.1$}&			291 600	&			 926	&   7.3	&	 7.1		&	1.53e-1	&	10005	\\
{2}&									&	{$\MF_{1.0\%}=2.2$}&			940 400	&			3014	&   2.3	&	 2.2		&	2.98e-2	&	10095\\
{3}&									&	{$\MF_{0.1\%}=1.1$}&		  1 848 700 &			6013	&   1.1	&	 1.1		&	2.81e-3	&	10125	\\
\midrule
{4}&\multirow{ 3}{*}{MASSPEED}			&	{$\MS_{5.0\%}=13\;049$}&			 16 801	&			  57	&	119	&	103		&	1.19e-2	&	10125	\\
{5}&									&	{$\MS_{1.0\%}=2\;985$}	&			 34 802	&		 114	&	 60	&	50		&	4.17e-3	&	10125	\\
{6}&									&	{$\MS_{0.1\%}=304$}	&			109 101	&			 353	&	 19	&	16		&	1.06e-3	&	10125	\\
\midrule
{7}&\multirow{ 4}{*}{A-MASSPEED}		&	{$\MS_{5.0\%}$} &			  2 236	&			   9	& 782	& 836		&	3.81e-1	&	10485	\\
{8}&									&	{$\MS_{1.0\%}$} &			  4 700	&			  16	& 413	& 404		&	9.36e-2	&	10215	\\
{9}&									&	{$\MS_{0.1\%}$} &			 14 234	&			  48	& 141	& 132		&	1.19e-2	&	10125	\\
{10}&									&	{$\MS_{0.01\%}$}&			 44 443	&			 150 	&  45	&  42		&	1.61e-3	&	10125	\\
\bottomrule
\end{tabular}
\label{tab:resumeAg005}
\end{center}
\end{table}
%---------------------------------------------------------------------%

For this test we performed 10 different runs summarized in Tab.~\ref{tab:resumeAg005}.
The approach used is specified in column 2, while the acceleration factors and the tollerance used to compute them is specified in the column.
Runs 1 to 3 implement the MORFAC approach ($\Mcw=\Mq=1$, and $\Mcs=\MF$), with tolerance $\Tol$ equal to 5\%, 1\% and 0.1\% respectively.
Runs 4 to 6 implement the MASSPEED approach ($\Mq=1$, and $\Mcw=\Mcs=\MS$), with the same tolerances as the previous set.
The constant factors $\MF$ and $\MS$ are computed by using \eqref{eq:linexp} and \eqref{eq:linexpMS}, respectively and considering the highest Froude number of the simulation.
In this test the maximum Froude number corresponds to the initial maximum value $\Fr\approx0.23$, as shown in Fig.~\ref{fig:BedFrEvo}b.
Finally, runs 7 to 10 implement the adaptive version A-MASSPEED, tested with tolerances $\Tol=5;1;0.1;0.01$ [\%]. 

\begin{figure}[tb]%
\centering
\includegraphics[width=\columnwidth]{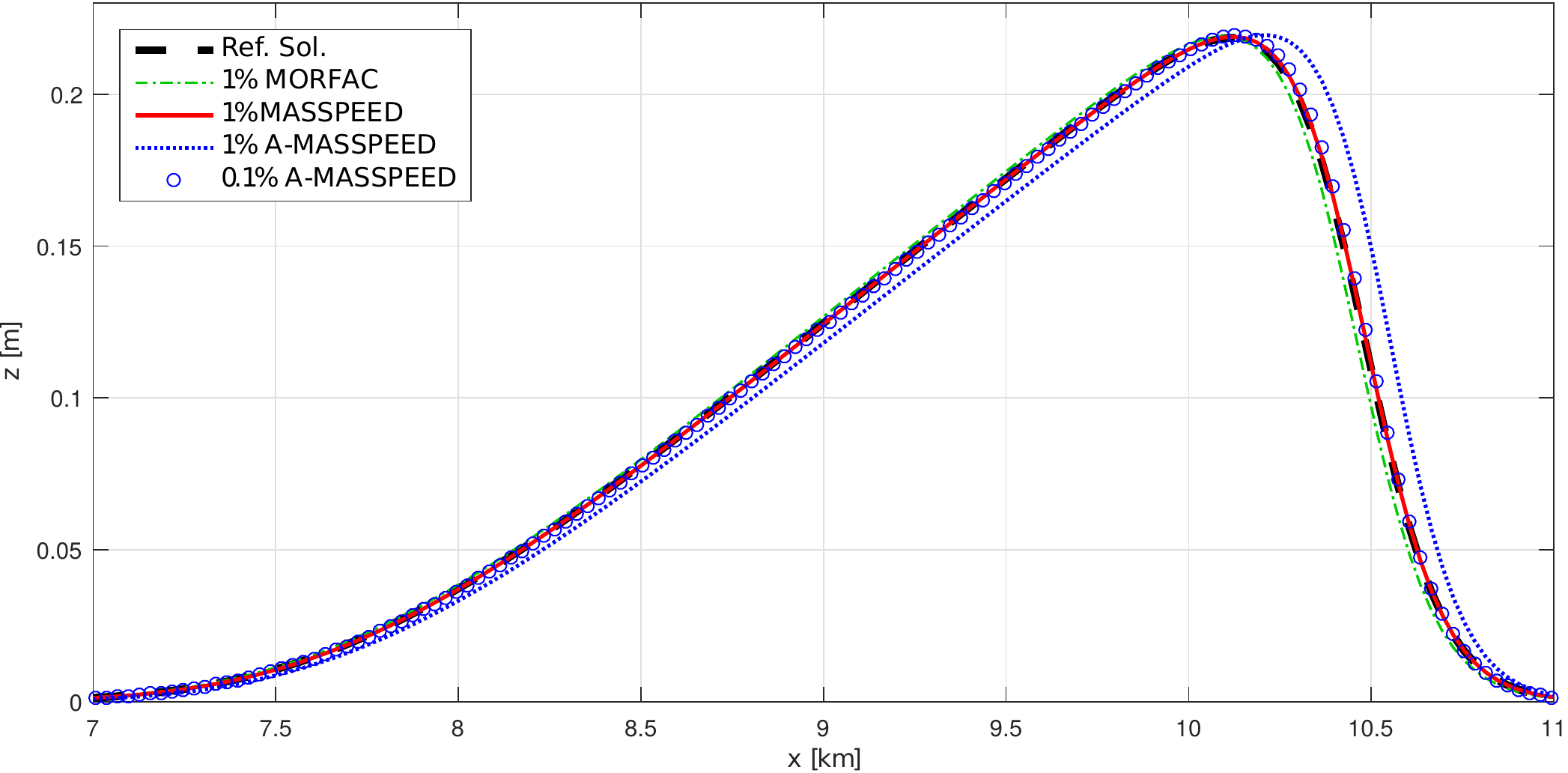}%
\caption{Bottom hump profiles after a 100 days evolution: comparison between reference solution (thick black), MORFAC (run 2), MASSPEED (run 5) and A-MASSPEED (runs 8,9) models.}%
\label{fig:HumpEND}%
\end{figure}

The accuracy of the accelerated solutions with respect to the reference solution is evaluated qualitatively in terms of the final position of the hump crest and, quantitatively, via the normalized root square error $\Eg$, defined as
\begin{equation}
\Eg = \frac{\sqrt{\sum{\left(z-z_\mathtt{ref}\right)^2}}}{\sqrt{\sum{z_{\mathtt{ref}}^2}}}\,,
\label{eq:Error}
\end{equation}
where $z$ and $z_\mathtt{ref}$ are the bottom profiles at the end of the simulation for the given accelerated model (MORFAC or MASSPEED) and the original system.
The computational costs and accuracy of the 10 runs are also specified in Tab.~\ref{tab:resumeAg005}, while some examples of final bottom profiles are depicted in  Fig.~\ref{fig:HumpEND}. 

As a first robustness assessment, it is worth noting that the accuracy ($\Eg$ and crest position) increases when reducing the tolerance $\Tol$ for all the acceleration strategies.
Table \ref{tab:resumeAg005} highlights also the significant differences of computational speed-up between the tested approaches: i.e. MORFAC and A-MASSPEED have speed-up differences of roughly one order or magnitude.
Moreover, it is important to underline that the theoretical speed-up $\Sp$ is always fairly close to the measured $\Sp^\mathtt{CPU}$ (see \S \ref{sec:def_SpSpCPU}). 
Hence the theoretical formulation can be adopted as a good \textit{a priori} estimation of the effective simulation speed-up.  

\begin{figure}[tb]%
\centering
\includegraphics[width=\columnwidth]{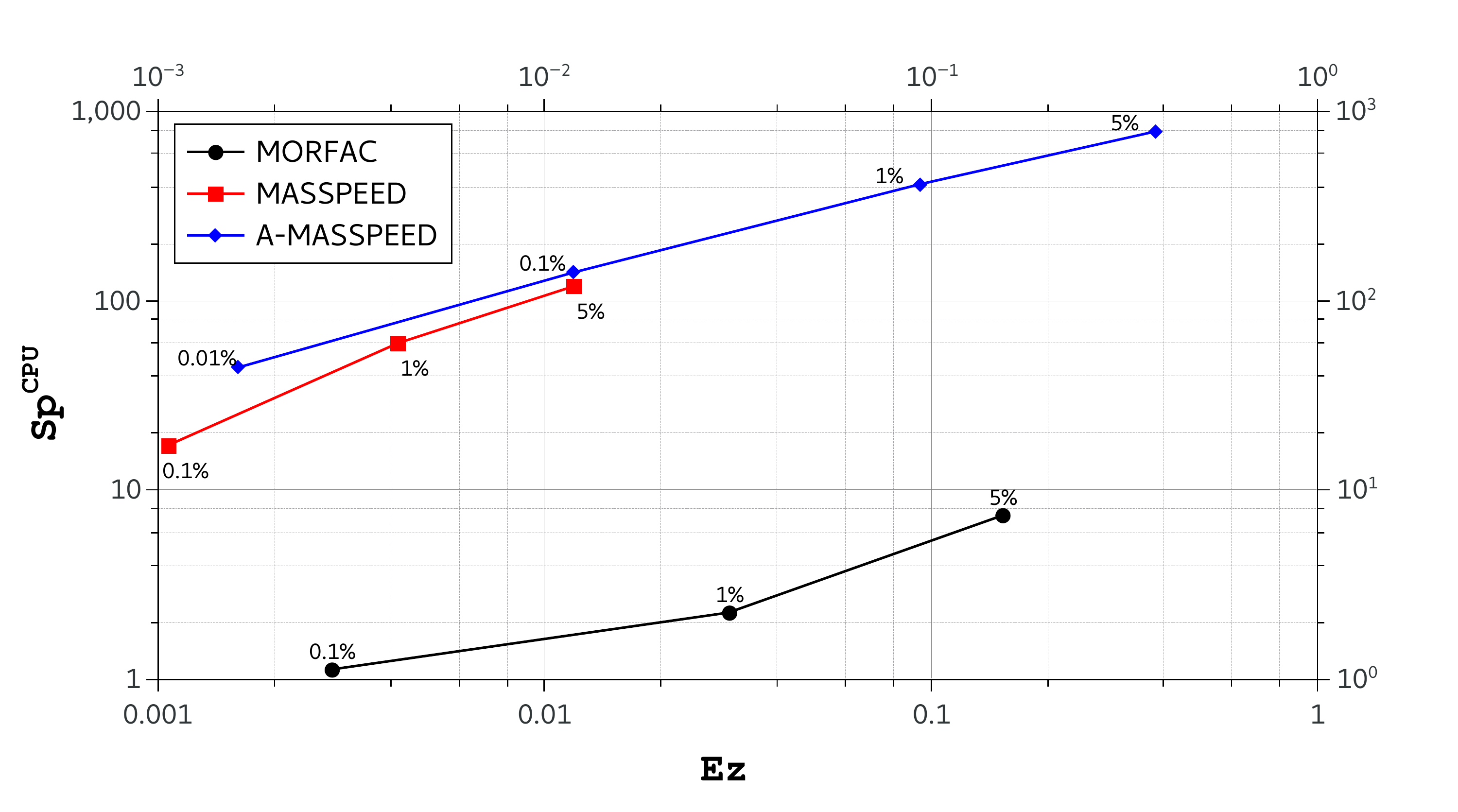}%
\caption{Numerical speed-up $\Sp^\mathtt{CPU}$ against normalized errors $\Eg$ for the 10 accelerated runs. Marker labels highlight the user-defined tolerances resumed in Table \ref{tab:resumeAg005}.}%
\label{fig:EzSp}%
\end{figure}

To better focus on the differences between the investigated strategies, let us discuss Fig.~\ref{fig:EzSp}, where the numerical speed-up $\Sp^\mathtt{CPU}$ of the 10 accelerated runs are plotted against the normalized errors $\Eg$.
It is worth noting that the plot is in log-log scale.
Beside the evident increase of the computational speed-up offered by the new proposed approaches with respect to MORFAC, Fig.~\ref{fig:EzSp} sheds the light on the accuracy differences of the three methods for a given $\Tol$. 
In particular, it appears that for the same given $\Tol$,  i) MASSPEED is more accurate than MORFAC and ii) the adaptive A-MASSPEED is less accurate than both fixed methods. 
To justify issue i), please note that for the MORFAC and MASSPEED methods we \textit{fix} the acceleration factors $\MF$ and $\MS$ (values in Tab.~\ref{tab:resumeAg005}) for the entire run based on the maximum expected Froude number and the given tolerance. 
If the Froude number decreases during the simulation, as in the given test, the error on the linearity assumption \eqref{eq:linexp} also decreses, being a function of $\Fr$ (Fig.~\ref{fig:MORSvsMORF}a and b). 
Nevertheless, the error reduction is faster for the MASSPEED approach, i.e. the ratio of the bottom eigenvalue (red solid line) in Fig.~\ref{fig:MORSvsMORF}b tends to the constant factor $\MS$ faster than the corresponding ratio in Fig.~\ref{fig:MORSvsMORF}a. 
Given this difference in the error reduction rate, the accumulated final error of the MASSPEED method is reduced with respect to MORFAC.    

Issue ii) well underlines the main difference between the \textit{fixed} approach (MORFAC and MASSPEED) and the \textit{adaptive} one (A-MASSPEED).
This accuracy gap is evident also in Fig.~\ref{fig:HumpEND}, where final bottom profiles are plotted for some selected runs.
For given $\Tol = 1$\%, MORFAC and MASSPEED profiles are very close to the reference solution, while the A-MASSPEED solution shows an offset of the crest position.
As previously discussed, the effective error on the linearity Eq.~\eqref{eq:linexp} decreases during the simulation time when using the \textit{fixed} approaches.
On the other hand, with A-MASSPEED, the \textit{adaptive} acceleration factor $\MS$ is not bonded by  $\Fr^{\max}$, but instead recomputed at each time step.
This means that the error on the linearity assumption \eqref{eq:linexp} is dynamically forced to be constant and equal to the user-defined $\Tol$.
Such different behavior of the error, decreasing with the fixed strategy but constant with the adaptive one, results in the final greater error of the latter.

Even if, for a given user-defined $\Tol$, the \textit{adaptive} strategy is less accurate, it it important to highlight that it globally outperforms both the \textit{fixed} methods: as a matter of fact, for any accuracy (error $\Eg$), \mbox{A-MASSPEED} provides the highest numerical speed-up (Fig.~\ref{fig:EzSp}). Moreover, this method has a further benefit related to the robustness. In fact with the \textit{fixed} approach, we need to known \emph{a priori} the maximum Froude number occurring during the simulation. For very simple applications, as in the test presented here, this prediction is straightforward but in case of more complex hydro-morphological configurations this might not be possible. A "wrong" initial setting of the constant acceleration factors may lead to a final numerical solution that did not satisfy the linearity conditions, hence to a non-linear acceleration of the morphological evolution. In the worst case, a wrong constant factor leads to the loss of hyperbolicity of the accelerated system, hence to a completely failing numerical solution. 
On the contrary, such troubles are inherently handled by the \textit{adaptive} strategy, where the acceleration factors are dynamically recomputed to keep the acceleration within the linearity threshold.

\subsection{Long term evolution of a bottom hump: bottom friction case}
\label{sub:friction}
The main goal of this test is to verify that the friction term does not alter the main features of the accelerated models.
To account also for the bottom friction term into the accelerated framework, system \eqref{eq:NUaccSystem} can be rewritten as
\begin{equation}
\frac{\partial \W}{\partial t}+\M\,\A(\W)\frac{\partial \W}{\partial x} = \M\bm{S}(\W),
\label{eq:qLinAcc+sf}
\end{equation}
where $\bm{S}(\W)$ is the vector of the source terms defined in \eqref{eq:def_WAS} and $\M$ is the acceleration matrix.
It is important to note that $\M\bm{S}(\W) = [ 0 ,	-\Mq\,c^2 s_f , 0 ]^\text{T}$ and $\Mq=1$ in both MORFAC and MASSPEED approaches.
Therefore, the accelerations of the models do not influence the expression of the friction source term.

From a numerical point of view, the source term is treated by using a classical splitting procedure (e.g. \cite{Siviglia2013}).
Without loss of generality, we present the results only for the adaptive A-MASSPEED approach, assuming $\Tol=0.001$. 

\begin{figure}[tb]%
	\centering
	\subfloat 
	{\includegraphics[width=0.48\columnwidth]{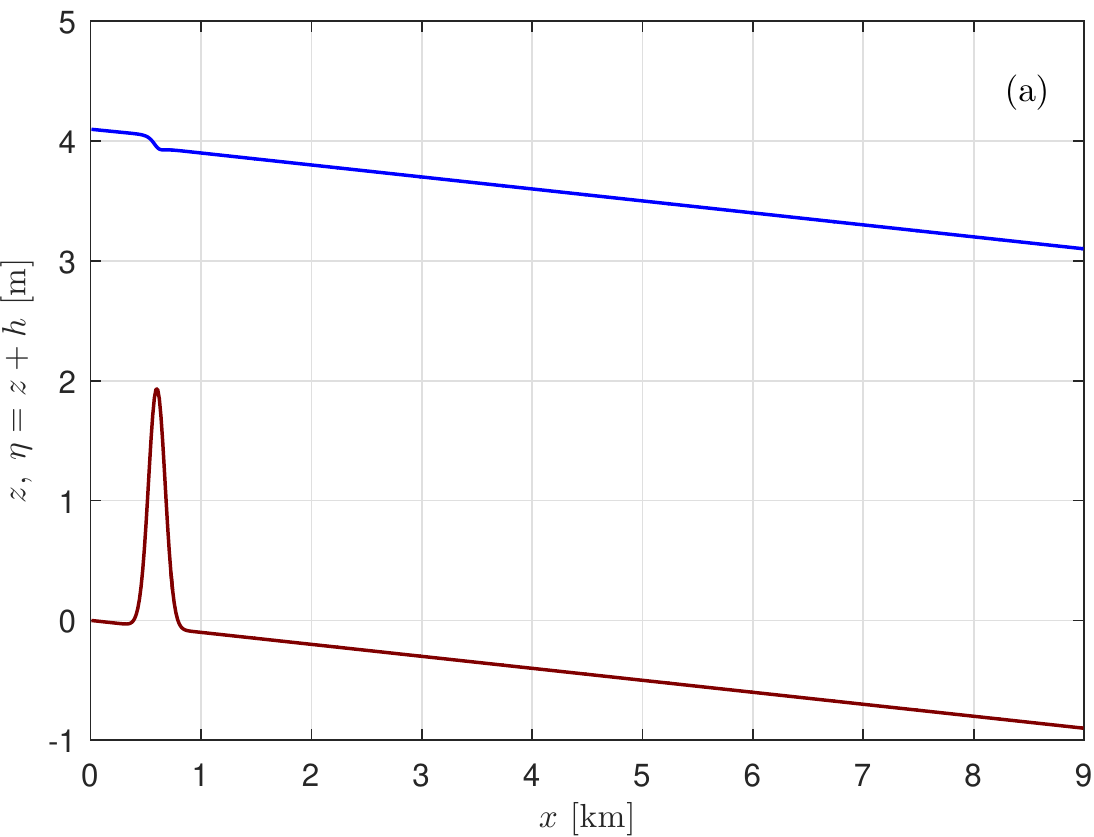}}\quad
	\subfloat 
	{\includegraphics[width=0.485\columnwidth]{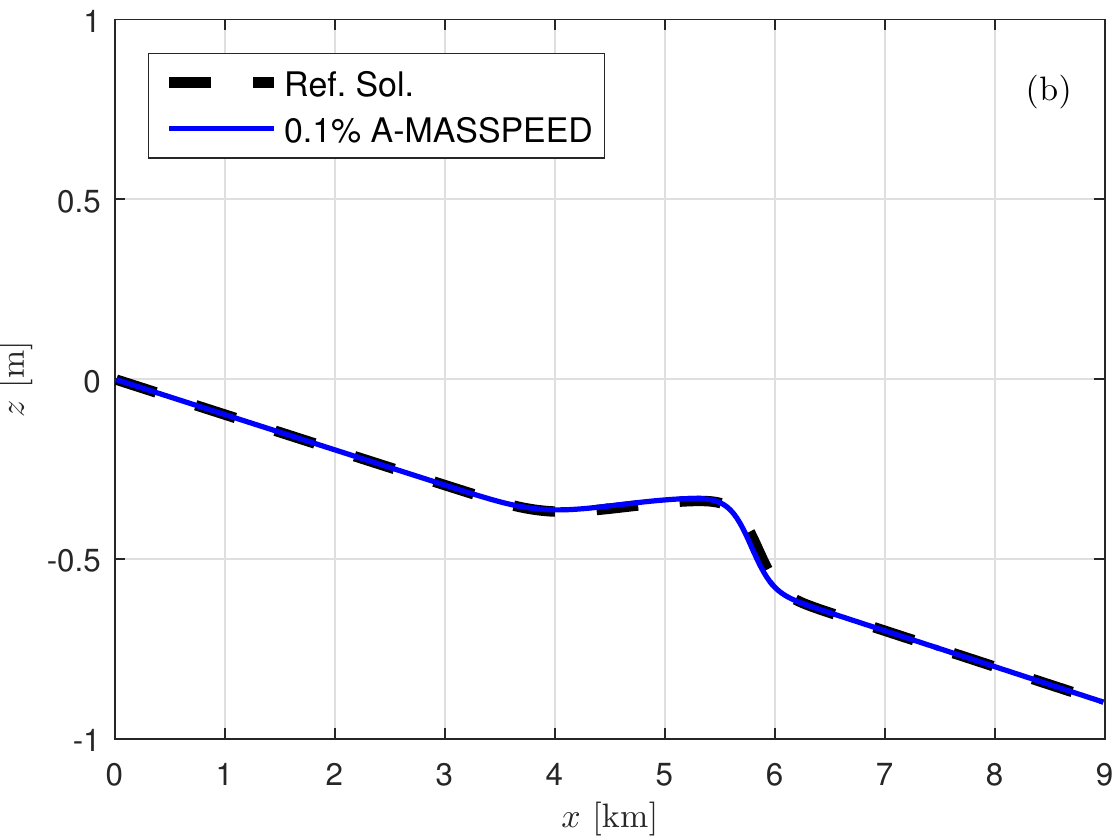}}\\
\caption{Treatment of the friction source term: (a) initial condition of the Hump test with not negligible friction; (b) results of a 50 days long morphodynamic simulation of the hump evolution.}%
\label{fig:FrictionHump}%
\end{figure}

The test is a modification of that given in the previous section (same boundary conditions), where we add a constant slope $s_0 = 0.1$\textperthousand~ to the initial bed profile, thus,
\begin{equation}
z(x,0)= - s_0 x + z_{\max}\, \exp\left[{-\frac{(x-\mu)^2}{\sigma^2}}\right].
\label{eq:HumpIC_FRX}
\end{equation}
Assuming a water discharge $q_0= 2$ m$^3$/(s m) and constant Strickler roughness $K_{s} = 19.8$ m$^{1/3}$s$^{-1}$, the resulting normal flow depth is $h_0=4$ m.
Fig.~\ref{fig:FrictionHump}a shows the initial water surface and bottom elevation profiles. The simulated time is 50 days.

%---------------------------------------------------------------------%
%---------------------------------------------------------------------%

Fig.~\ref{fig:FrictionHump}b shows the comparison between the bottom profile at the end of the simulation, obtained using the A-MASSPEED approach, and the corresponding reference solution.
It is seen that the solutions are in good agreement.
Moreover, the error $\Eg$ computed by using relation \eqref{eq:Error}, is equal to 1.3e-2, i.e., of the same order of magnitude of the errors obtained by neglecting the friction term (see Tab.~\ref{tab:resumeAg005}).
Analogous results, here omitted for the sake of brevity, can be obtained by using the MORFAC approach. 

\section{Conclusions and future trends}
\label{sec:conclusion}
In this paper we carried out a mathematical study to identify the conditions under which bed evolution, governed by the one-dimensional dSVE equations, can be linearly accelerated.
This was achieved by introducing the concept of non-uniform acceleration, consisting by multiplying the spatial derivatives of each of the three governing equations by a constant acceleration factor.
Then, from the  study of a simplified linear solution of the eigenvalues of the non-uniformly accelerated system  we  
\begin{enumerate}
\item show that the classical MORFAC acceleration results from the acceleration of the mere sediment continuity equation;
\item obtain MASSPEED, a new linear morphodynamic acceleration technique, in which both mass continuity equations for water and sediment are accelerated by the same quantity;
\item set the basis for the derivation of a  criterion for the a priori determination of the   highest acceleration factor. It turned out that the MASSPEED can be successfully applied on a wider range of flow and sediment transport conditions as compared with the classical MORFAC approach;   
\end{enumerate}
  
Building on the knowledge obtained from the linear analysis,  we derived a practical and implementable criterion for the determination of the maximum acceleration factor for both techniques.
The accuracy of the numerical solution can be determined by the user through the choice of a small tolerance value.
This quantifies the extension of the range of validity under which the bed level can be linearly accelerated.
The new criterion was implemented within an existing code following an adaptive concept in a similar manner as the CFL stability condition. 
Thanks to this adaptive procedure, the  maximum accelerating factor is chosen at each time step according to the actual flow and sediment transport conditions. 
The numerical solution of the long term evolution of a sediment hump demonstrated that the application of the MASSPEED approach results in larger speed-up and considerable reduction  of the computational time.
It is also worth remarking the most important limitations of the proposed approach.
The results presented  are based on a few, strategic model simulations of the simple case of the one-dimensional propagation of a sediment hump under almost constant flow conditions.
Therefore, they must be cautiously used when non-uniform morphology and time varying nonlinear forcing which may include tides, flood waves, etc. comes to modelling.
It is likely that, in these cases, the maximum acceleration factor must be considerably decreased. 
In particular, we found that the higher the $\Fr$ is, the smaller will be the acceleration; furthermore, it is likely that the more unsteady the flow is, the smaller will be the possible acceleration.
Bearing in mind these limitations, the theoretical background presented in this provides a robust basis for further exploring new morphological accelerators and extending the limit of applicability of the MASSPEED approach. 
Moreover, following a similar approach, the application of the MASSPEED approach can be extended to future trends, including:
unsteady flow conditions, 2D (planar) morphological models, suspended transport and the de Saint Venant-Exner-Hirano model for non-uniform sediment deposition.

\section*{Acknowledgements} 
Part of this work has been carried out during Francesco Carraro's visit to the Laboratory of Hydrology, Hydraulics and Glaciology of the Swiss Federal Institute of Technology. His staying was partially supported by the University of Ferrara through the 5\textperthousand~ fees donation within the ``Young Researcher Project''.
Valerio Caleffi's research has been funded by the University of Ferrara within the Founding Program FIR 2016, project title ``Energy-preserving numerical models for the Shallow Water Equations''.
We are also grateful to Prof. K. Hutter for reading the first draft of the manuscript.

\appendix
\section{Analytical study of the the bottom evolution}
\label{app:Val}
% !TEX root = 2_MASSPEED.tex
%\newcommand{\UC}{\bm{U}}     			% val
%\newcommand{\LL}{\bm{L}}				% val
%\newcommand{\RR}{\bm{R}}				% val
%
For $\Fr \ll 1$ and $\psi \ll 1$, the governing system \eqref{eq:def_SW-Ex} is well approximated by the \emph{decoupled form} \eqref{eq:def_SW-Ex_decoupled}. From a mathematical point of view, Eq.~\eqref{eq:def_SW-Ex_decoupled} implies that the decoupling is possible when the conservative variable $z$ coincides with one of the three characteristic variables of the morphodynamic problem. A worthy example to better understand this feature is the linearised problem proposed in \cite{Lyn2002} and considered in \S \ref{sec:MathSol}.

To verify the validity of the decoupled formulation, the contribution of each eigenvalue (and the corresponding eigenvector) to the bed evolution is analysed, both in terms of bed forms celerity and amplitude.

The prototype problem  \cite{Lyn2002} is governed by the \emph{linearized original system} of Eqs.~\eqref{eq:LinearSystem}-\eqref{eq:linear_A} (here reproduced for readability),
\begin{equation}
	\frac{\partial\W}{\partial t}+\A_\U\frac{\partial\W}{\partial x} = 0
	\quad\text{with:}\quad
	\A_\U=\A(\W_\U)=
	\begin{bmatrix}
		0                   &	1	        &	0       \\
		c_\U^2-u_\U^2	        &   2u_\U    &	c_\U^2		\\
		-u_\U\,\psi_\U 		&	\psi_\U		&	0
	\end{bmatrix}	\,,
\label{eq:LinearSystem_Apx}
\end{equation}
where subscript $\U$ refers to the unperturbed state used to linearise the problem.
Therefore, given the vector of unperturbed conservative variables $\W_\U=[h_\U,\,q_\U,\, z_\U]^T$, 
$u_\U=q_\U/h_\U$ is the flow velocity;
$c_\U=\sqrt{g\,h_\U}$ is the unperturbed celerity;
while $\psi_\U$ is the uniform sediment transport as defined in \eqref{eq:phi_grass}.

The initial condition of the problem is given by
\begin{equation}
\W_0(x)= \begin{bmatrix} h_\U \\	q_\U \\ z_0 \end{bmatrix}\,,
\end{equation}
where: $z_0(x) = z_{\max} \exp(-x^2)$ with $z_{\max}=1.0^{-5}$ m, $h_\U=1$ m, $q_\U = \sqrt{h_\U^3\,g\,\mathtt{Fr}_\U^2\rule{0ex}{2.1ex}}$, $\mathtt{Fr}_\U=0.7$ and $\psi_\U = 0.01$. Finally, a propagation time $t_p=15$ s is assumed.

The linearized system \eqref{eq:LinearSystem_Apx} can be analytically solved by adopting the characteristic method \cite{Toro2009} and using the analytical eigenvalues and eigenvectors of \ref{apx_A} (with $\Mcw = \Mq = \Mcs = 1$).
Thus, the conservative variables $\W(x)$ can be projected on the characteristic space multiplying them by the inverse of the matrix of the right eigenvectors of $\A_\U$,
\begin{equation}
	\UC(x)=\LL_\U\,\W(x)\,,
	\label{eqV:lin_sol_dir}
\end{equation}
with $\UC(x)$ the vector of the characteristic variables and $\LL_\U=\RR_\U^{-1}$ the inverse of the matrix of the right eigenvectors computed for the unperturbed state.
Therefore, the initial conditions in terms of characteristic variables are $\UC_0(x)=\LL_\U\,\W_0(x)$.

Each characteristic variable $\UC^{(j)}$ satisfy a linear advection equation with a celerity given by the corresponding constant eigenvalue, $\lambda_j^\U$, of $\A_\U$, i.e.,
\begin{equation}
\frac{\partial \UC^{(j)}}{\partial t}+\lambda_j^\U\frac{\partial \UC^{(j)}}{\partial x} = 0\quad \text{for} \quad j=1,\,2,\,3\;.
\end{equation} 
The corresponding solution in terms of $\UC$, for a given $x$ and $t$ is
\begin{equation}\label{eq:linconv}
\UC^{(j)}(x,t)=\UC_0^{(j)}(x-\lambda_j^\U\,t) \quad \text{for} \quad j=1,\,2,\,3\;.
\end{equation}
The evolved conservative variables $\W(x,t)$ can be found by multiplying $\UC(x,t)$ by the matrix of the right eigenvectors $\RR_\U$,
\begin{equation}
\W(x,t)=\RR_\U\,\UC(x,t). \label{eqV:RU}
\end{equation}

Focusing the attention on the bottom evolution, and indicating with $\mathbf{r}_j^{(3)}$ the third component of the $j$-th eigenvector of $\A_\U$, the third equation of the system \eqref{eqV:RU} can be written as
\begin{equation}
z(x,t)=\sum_{j=1}^3  \mathbf{r}_j^{(3)}\,\UC^{(j)}(x,t) = 
\sum_{j=1}^3  \zeta_j(x,t)\;, \label{eqV:RUz}
\end{equation}
where $\zeta_j(x,t)=\mathbf{r}_j^{(3)}\,\UC^{(j)}(x,t)$ represents the contribution of each component of the characteristic variables to the bottom evolution. 
%-----------------------------------------------%
\begin{figure}[tp]%
	\centering
	\includegraphics[width=0.95\columnwidth]{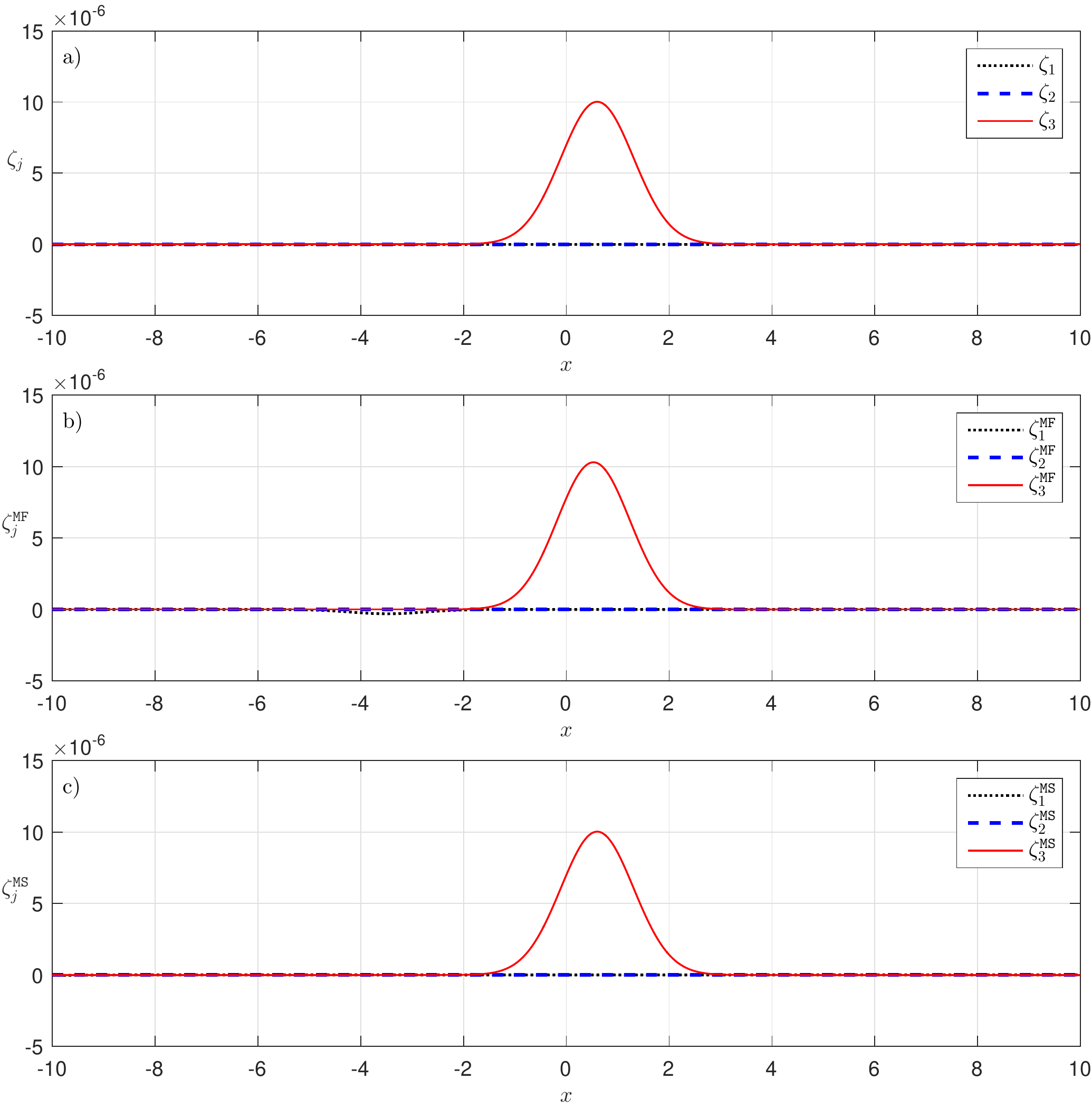}%
	\caption{The contributions of the three characteristic variables $\zeta_j$ to the bottom topography ($\mathtt{Fr}_\U=0.7;\;\psi_\U=0.01$). a) solution of the linearized original system at $t_p$; b) $\zeta_j^\MF$ solution of the linearized MORFAC system with $\MF = 5$ at $t_p/\MF$; c) $\zeta_j^\MS$ solution of the linearized MASSPEED system with $\MS = 5$ at $t_p/\MS$.}%
	\label{fig:contribFondo}%
\end{figure}
%-----------------------------------------------%

Numbering the eigenvalues as defined in \S\ref{sec:approx_sol} and computing the eigenvectors as in Eq.~\eqref{eq:RR}, Fig.~\ref{fig:contribFondo}a shows the comparison of the three terms $\zeta_j(x,t)$.
The figure shows that the only meaningful contributio to the bed evolution is given by $\zeta_3(x,t)=\mathbf{r}_3^{(3)}\,\UC^{(3)}(x,t)$ that implies $\zeta_3(x,t)\approx z(x,t)$.
Furthermore, because the time evolution of $\UC^{(3)}(x,t)$ is governed by the linear advection equation \eqref{eq:linconv}, the bed form migration celerity is $\lambda_3^\U = \lambda_b$, as stated by Eq.~\eqref{eq:wave_eq}.\bigskip

Appling again the characteristic method \cite{Toro2009} to the linearized MORFAC and MASSPEED systems (see, \eqref{eq:linear_A} of \S\ref{sec:MathSol}), accelerated dSVE problems can be solved.
According to \eqref{eq:linconv} and \eqref{eqV:RUz}, the bottom evolution is properly accelerated if the characteristic variable $\zeta_3 \approx z$ is not altered by the acceleration, while the corresponding eigenvalues $\lambda_3^\U$ increase proportionally to the acceleration itself.
For example, if accelerations of $\MF$ and $\MS$ are applied, the bed evolution is well represented if, at the \emph{scaled propagation times} $t_p^\MF=t_p/\MF$ and $t_p^\MS=t_p/\MS$, the accelerated bottom elevations can be expressed as
\begin{equation}
	\zeta_3^\MF(x,t_p^\MF) \approx \zeta_3^\MS(x,t_p^\MS) \approx \zeta_3(x,t_p) \approx z(x,t_p),
\label{eq:Ex_zita}
\end{equation}
with the accelerated eigenvalues
\begin{equation}
\frac{\lambda_3^\MF}{\MF} \approx \frac{\lambda_3^\MS}{\MS} \approx \lambda_3^\U\,.
\label{eq:Ex_lamda}
\end{equation}
%-----------------------------------------------%
\begin{figure}[tp]%
\centering
	\includegraphics[width=0.95\columnwidth]{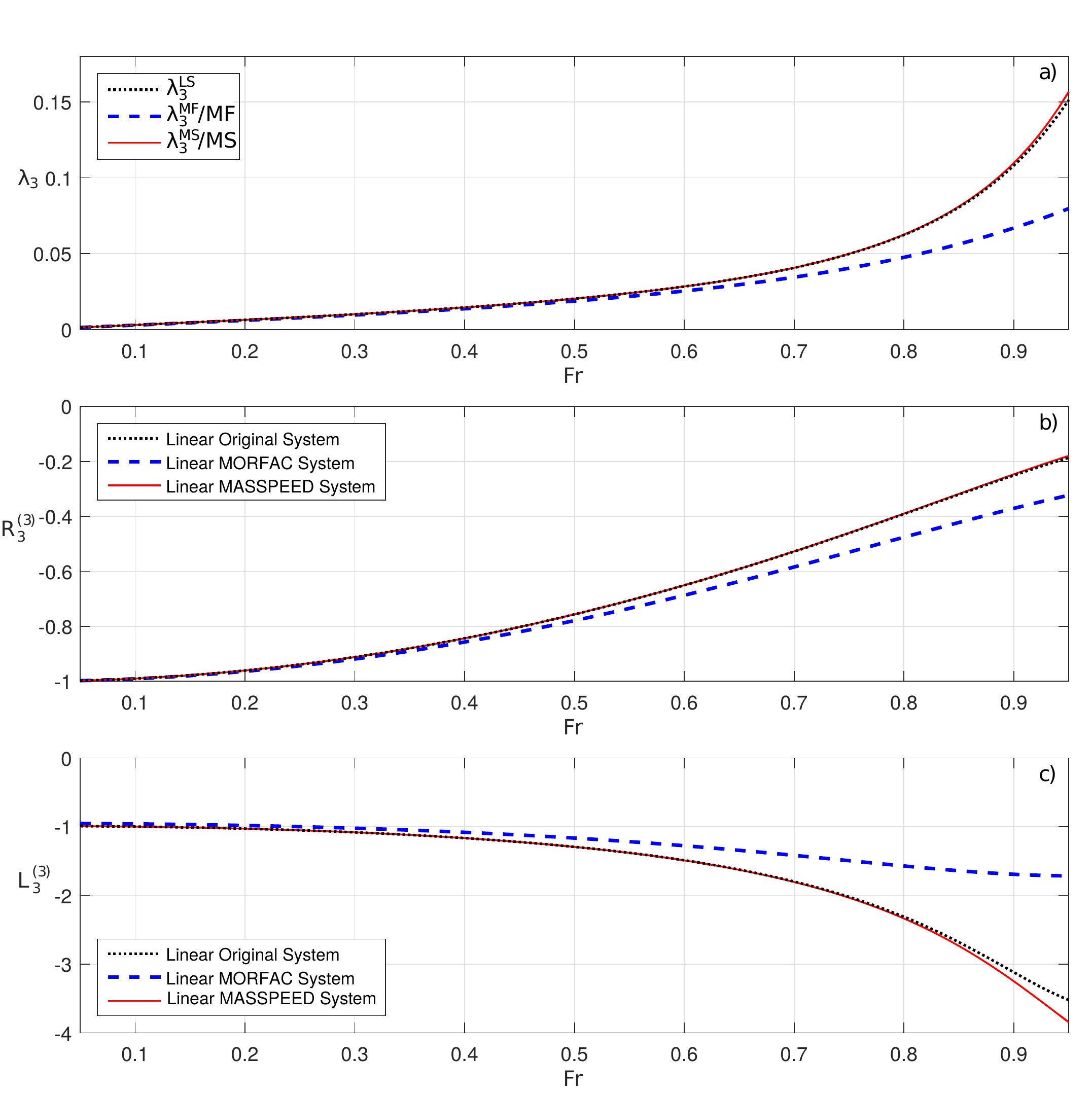}%
	\caption{Comparison between eigenstructure components for $0<\mathtt{Fr}_\U<0.95$, $\psi_\U = 0.01$ and $\MF=\MS=5$: a) acceleration of the third eigenvalue according to Eq.~\eqref{eq:Ex_lamda}; b) $\mathbf{r}_3^{(3)}$ according to the linearised original system, the linearised MORFAC system and the linearised MASSPEED system; c) $\mathbf{l}_3^{(3)}$ according to the linearised original system, the linearised MORFAC system and the linearised MASSPEED system.}%
	\label{fig:R32comp}%
\end{figure}
%-----------------------------------------------%
For this linearised example with $\MF=\MS=5$ and $t_p^\MF=t_p^\MS=3$ s, Eq.~\eqref{eq:Ex_zita} is verified, as shown in Fig.~\ref{fig:contribFondo}b (MORFAC system) and Fig.~\ref{fig:contribFondo}c (MASSPEED system).
Indeed, according to Eq.~\eqref{eqV:lin_sol_dir} and \eqref{eqV:RUz}, the evolved bottom function is related to both $\mathbf{r}_3^{(3)}$ and $\mathbf{l}_3^{(3)}$, so that any acceleration strategy must not alter these two quantities.
The component $\mathbf{r}_3^{(3)}$ of the three linear systems, from Eqs.~\eqref{eq:phi_grass} and \eqref{eq:RR} can be written as
\begin{equation}
	\mathbf{r}_3^{(3)}= \psi\,\frac{\Mcs}{\Mcw}\left(1-\frac{u\,\Mcw}{\lambda_3} \right)\,,
	\label{eq:RR_psi}
\end{equation}
which becomes
\begin{align}
	\mathbf{r}_3^{(3)}\left|_\mathtt{OS}\right. &= \psi\left(1-\frac{u}{\lambda^\U_3} \right)\,,\quad &&\text{for the original system,} \label{eq:R3LOS}\\
	\mathbf{r}_3^{(3)}\left|_\MF\right. &= \psi\,\MF\left(1-\frac{u}{\lambda^\MF_3} \right)\,, \quad &&\text{for the MORFAC system,} \label{eq:R3LMF}\\
	\mathbf{r}_3^{(3)}\left|_\MS\right. &= \psi\left(1-\frac{u\,\MS}{\lambda^\MS_3} \right)\,, \quad &&\text{for the MASSPEED system.} \label{eq:R3LMS}
\end{align}

For the MASSPEED approach, looking at Eqs.~\eqref{eq:R3LOS} and \eqref{eq:R3LMS}, it is trivial to see that, if condition \eqref{eq:Ex_lamda} is verified and then $\mathbf{r}_3^{(3)}\left|_\MS\right.\approx\mathbf{r}_3^{(3)}\left|_\mathtt{OS}\right.$, the acceleration technique does not alter the aplitude of the bed form.
Conversely, for the MORFAC approach, looking at Eqs.~\eqref{eq:R3LOS} and \eqref{eq:R3LMF}, condition \eqref{eq:Ex_lamda} is necessary but not sufficient to have $\mathbf{r}_3^{(3)}\left|_\MF\right.\approx\mathbf{r}_3^{(3)}\left|_\mathtt{OS}\right.$.
Due to the lack of acceleration of the hydrodynamic continuity equation, MORFAC approximate $\mathbf{r}_3^{(3)}$ well only if $\MF \approx 1$.

We can conclude that an exact linear scaling of the third eigenvalue is crucial to reproduce the propagation time of a bottom hump well, but also to preserve the correct profile of the riverbed.
As discussed in \S\ref{sec:LmaT}, the MASSPEED extends the range of linear scaling of $\lambda_3^\MS/\MS$ with respect to $\lambda_3^\MF/\MF$ of the MORFAC approach.

This general result is confirmed also by Fig.~\ref{fig:R32comp}: it shows, for $0<\Fr_\U<0.95$, the comparison between the third eigenvalue and the corresponding component of $\RR$ and $\LL$.
According to Fig.~\ref{fig:R32comp}a, assuming as reference the original system, the MASSPEED gives a better approximation of the third eigenvalue, specially for $\Fr>0.6$ (for a more extended analysis of the effects of the parameter changes on the $\lambda_3$ eigenvalue, see \S\ref{sec:Non_Lin_An}).
As a consequence, with respect to the classical MORFAC approach, the MASSPEED leads also to a much better approximation of $\mathbf{r}_3^{(3)}$ (Fig.~\ref{fig:R32comp}b) and $\mathbf{l}_3^{(3)}$ (Fig.~\ref{fig:R32comp}c).
Thus, the improvement obtained by using the MASSPEED approach with respect to the MORFAC approach is clear.

\section{Analytical eigenvalues and eigenvectors of the general governing system}
\label{apx_A}
% !TEX root = 2_MASSPEED.tex
%\newcommand{\UC}{\bm{U}}     			% val
%\newcommand{\LL}{\bm{L}}				% val
%\newcommand{\RR}{\bm{R}}				% val
%
The analysis presented in this work is based on the eigenvalues and eigenvectors of different governing systems, e.g. the original system, Eq.~\eqref{eq:qLin}, or the accelerated system, Eq.~\eqref{eq:NUaccSystem}. Therefore, the analytical closed expressions of the eigenvalues and eigenvectors are useful.

In this appendix we give the explicit formulations of such eigenvalues and eigenvectors for the following very general system:
\begin{equation}
\frac{\partial \W}{\partial t}+\M\,\A\frac{\partial \W}{\partial x} = 0\,,
\label{eq:ApxNUaccSystem}
\end{equation}
where
\begin{equation}
	\W= \begin{bmatrix} h \\	q \\ z \end{bmatrix},  \quad
	\A(\W)=
	\begin{bmatrix}
0 &	1 & 0 \\
c^2-u^2	& 2u & c^2 \\
\xi\frac{\partial q_s}{\partial h} 	& \xi\frac{\partial q_s}{\partial q} &	0 \end{bmatrix},
\quad
\M = \begin{bmatrix}
\Mcw &	0    & 0    \\
0	 &	\Mq  & 0    \\
0	 &	0	 & \Mcs \end{bmatrix}\,.
	\label{eq:Apxdef_WAS}
\end{equation}

The characteristic polynomial of the system is obtained by evaluating$|\M\,\A(\W)-\lambda \bm{I}|=0$, i.e..
\begin{equation}
	\lambda^3-2\,\Mq\,u\,\lambda^2+\Mq\left(\Mcw\,\Fr^2-\Mq - \Mcs\,\xi\frac{\partial q_s}{\partial q} \right)\,\frac{u^2}{\Fr^2}\,\lambda-\Mcs  \Mcw \Mq \frac{u^2}{\Fr^2}\,\xi\frac{\partial q_s}{\partial h}=0\,,
	\label{eq:charPol}
\end{equation}
and the analytical solutions of the characteristic polynomial \eqref{eq:charPol} can be computed by using the Cardano formulas \cite{cardano} leading to
\begin{equation}
	\begin{aligned}
		\frac{\lambda_{1}}{c}&=\frac{2}{3}\Mq\,\Fr-\frac{2}{3}\sqrt{k_{2}}\,\cos{\left(\frac{\phi}{3}-\frac{\pi}{3}\right)}		\,, \\
		\frac{\lambda_{2}}{c}&=\frac{2}{3}\Mq\,\Fr+\frac{2}{3}\sqrt{k_{2}}\,\cos{\left(\frac{\phi}{3}\right)}									\,, \\
		\frac{\lambda_{3}}{c}&=\frac{2}{3}\Mq\,\Fr-\frac{2}{3}\sqrt{k_{2}}\,\cos{\left(\frac{\phi}{3}+\frac{\pi}{3}\right)}		\,, \\
	\end{aligned}
	\label{eq:apxlambda_MASSPEED}
\end{equation}
where:
\begin{equation}
	\begin{aligned}
		&\phi=\arccos{\left(\frac{k_{1}}{\sqrt{4\,{k_{2}}^3}}\right)},\\
		&k_{1} = 2\,\Mq^2\,\Fr\left(8\,\Mq\,\Fr^2+9\,\Mcs\,\xi\frac{\partial q_s}{\partial q}+9\,\Mcw\left(1-\mathtt{Fr}^2\right)\right)+27\,\Mcw\,\Mq\,\Mcs\,\xi\frac{\partial q_s}{\partial h},	\\
		&k_{2} = 4\,\Mq^2\,\Fr^2+3\,\Mq\,\Mcs\,\xi\frac{\partial q_s}{\partial q}+3\,\Mcw\,\Mq\left(1-\Fr^2\right).
	\end{aligned}
\label{eq:apxMASSPEED_parameters}
\end{equation}
Given the eigenvalue $\lambda_i$, the associated right eigenvector $\mathtt{\mathbf{r}}_i$ can be computed by solving the linear system $\M\,\A\,\mathtt{\mathbf{r}}_i = \lambda_i\,\mathtt{\mathbf{r}}_i$.
Therefore, the matrix $\RR$ of the right eigenvectors can be expressed as a function of the eigenvalues, the result being
\begin{equation}
	\RR=\left[ \mathtt{\mathbf{r}}_1, \mathtt{\mathbf{r}}_2, \mathtt{\mathbf{r}}_3 \right]
		,\quad \text{with:}\quad
		\mathtt{\mathbf{r}}_i = \left[1,\; \frac{\lambda_i}{\Mcw},\;\xi\frac{\Mcs}{\Mcw}\left(\frac{\Mcw}{\lambda_i}\frac{\partial q_s}{\partial h}+\frac{\partial q_s}{\partial q}\right) 	\right]^\mathtt{T}.
	\label{eq:RR}
\end{equation}

For a given sediment transport formula that allows evaluating the $q_s$ derivatives, eigenvalues and eigenvectors can be computed by substituting in Eqs.~\eqref{eq:apxlambda_MASSPEED}, \eqref{eq:apxMASSPEED_parameters} and \eqref{eq:RR}, respectively:
$\left(\Mcw=\Mq=\Mcs=1\right)$ for the original system;
$\left(\Mcw=\Mq=1,\;\Mcs=\MF\right)$ for the MORFAC system;
and $\left(\Mq=1,\;\Mcw=\Mcs=\MS\right)$ for the MASSPEED system.


%\section*{References}
%\bibliographystyle{elsarticle-harv}
%\bibliography{ms}


\end{document}